\documentclass[a4paper,UKenglish,cleveref, autoref, thm-restate,authorcolumns]{lipics-v2019}
\usepackage[utf8]{inputenc}
\usepackage{algpseudocode}
\usepackage{algorithm}
\usepackage{algpseudocode}
\usepackage{blindtext}
\usepackage{hyperref}
\usepackage{tikz}
\tikzstyle{vertex}=[circle, draw, inner sep=0pt, minimum size=5pt]

\usetikzlibrary{decorations.markings}
\usetikzlibrary{decorations.pathreplacing}
\usetikzlibrary{arrows.meta}
\usepackage{anyfontsize}
\usepackage{graphicx}
\usepackage{amsmath}
\usepackage{amsfonts}
\usepackage{xcolor}
\usepackage{amssymb}
\newcommand{\boundellipse}[3]
{(#1) ellipse (#2 and #3)
}
\usetikzlibrary{shapes.geometric}
\tikzstyle{square}=[draw, shape=regular polygon, regular polygon sides=4,draw,inner sep=0pt,minimum
size=0.225cm]
\tikzstyle{triangle}=[draw, shape=regular polygon, regular polygon sides=3,draw,inner sep=0pt,minimum
size=0.3cm]

\definecolor{azure}{rgb}{0.0, 0.5, 1.0}
\definecolor{pink}{rgb}{0.84, 0.09, 0.41}
\definecolor{magenta}{rgb}{0.8, 0.0, 0.8}
\definecolor{cyan}{rgb}{0.0, 0, 1.0}
\definecolor{green1}{rgb}{0, 1, 0}
\definecolor{green}{rgb}{0, 1, 0}
\definecolor{grey}{gray}{0.5}
\definecolor{brown}{rgb}{0.65, 0.16, 0.16}
\definecolor{aquamarine}{rgb}{0.5, 1.0, 0.83}
\definecolor{battleshipgrey}{rgb}{0.52, 0.52, 0.51}
\definecolor{cadetgrey}{rgb}{0.57, 0.64, 0.69}

\newcommand{\claimqed}{\hfill $\lhd$}

\newtheorem{ucivs}{\bf Reduction Rule UCIVS}
\newtheorem{ucevs}{\bf Reduction Rule UCEVS}
\newtheorem{observation}{\bf Observation}
\newtheorem{evd}{\bf Reduction Rule UCVD}
\newtheorem{evd*}{\bf Reduction UCVD^*}
\newtheorem{eed}{\bf Reduction Rule UCED}
\newtheorem{eee}{\bf Reduction Rule UCEE}
\newtheorem{eea}{\bf Reduction Rule UCEA}
\newtheorem{disevd}{\bf Reduction Rule Disjoint-UCVD}

\usepackage{algorithm}
\usepackage{algpseudocode}
\usepackage{tcolorbox}

\title{Parameterized Algorithms for Editing to Uniform Cluster Graph} 

\titlerunning{Uniform Cluster Graph Modifications} 

\author{Ajinkya Gaikwad}{Indian Institute of Science Education and Research, Pune, India} {ajinkya.gaikwad@students.iiserpune.ac.in}{}{}
\author{Hitendra Kumar}{Indian Institute of Science Education and Research, Pune, India} {sahu.hitendrakumar@students.iiserpune.ac.in}{}{}
\author{Soumen Maity}{Indian Institute of Science Education and Research, Pune, India}{soumen@iiserpune.ac.in}{}{}

\authorrunning{A. Gaikwad, H. Kumar and S. Maity} 

\Copyright{} 

\category{} 

\relatedversion{} 

\supplement{}

\EventEditors{John Q. Open and Joan R. Access}
\EventNoEds{2}
\EventLongTitle{42nd Conference on Very Important Topics (CVIT 2016)}
\EventShortTitle{CVIT 2016}
\EventAcronym{CVIT}
\EventYear{2016}
\EventDate{December 24--27, 2016}
\EventLocation{Little Whinging, United Kingdom}
\EventLogo{}
\SeriesVolume{42}
\ArticleNo{23}
\nolinenumbers
\begin{document}

\maketitle

\begin{abstract}

This paper investigates the parameterized complexity of transforming graphs into Uniform Cluster graphs, where every component is an equal-sized clique. We consider a number of different graph editing operations. 
We investigate several variants of this problem, including \textsc{Uniform Cluster Vertex Deletion (UCVD)}, \textsc{Uniform Cluster Edge Deletion (UCED)}, \textsc{Uniform Cluster Edge Addition (UCEA)},  \textsc{Uniform Cluster Edge Editing (UCEE)}, \textsc{Uniform Cluster Exclusive Vertex Splitting (UCEVS)} and 
\textsc{Uniform Cluster Inclusive Vertex Splitting (UCIVS)}.
For {\sc UCVD}, we provide a vertex kernel of size $\mathcal{O}(k^{3})$ and an FPT algorithm with running time $2^{k} \cdot n^{\mathcal{O}(1)}$ which improves the $3^{k} \cdot n^{\mathcal{O}(1)}$ time algorithm in the literature.
For the edge based variants, we provide a vertex kernel of size $\mathcal{O}(k^{2})$ for {\sc UCEE}.
We also provide a linear vertex kernel of size $6k$ and $5k$ for {\sc UCED} and {\sc UCEA}, respectively. 
This improves the quadratic size vertex kernel for {\sc UCEA} in the literature. 
Apart from kernelization, we provide a $1.47^{k} \cdot n^{\mathcal{O}(1)}$ for {\sc UCED} which improves the $2^{k} \cdot n^{\mathcal{O}(1)}$ time algorithm in the literature. 
This algorithm is heavily based on the algorithm by B$\ddot{o}$cker and Damaschke [IPL 2011] for {\sc Cluster Edge Deletion}.
These results improve the results as well as answer multiple open questions by Neeldhara Misra, Harshil Mittal, Saket Saurabh and Dhara Thakkar [ISSAC 2023].

Studying clustering problems on dense graphs is vital as they often allow for specialized algorithms that exploit the high connectivity, leading to significantly improved time complexities. 
For example, while clustering on general graphs may require single-exponential algorithms, dense graphs may enable sub-exponential solutions, highlighting the theoretical and practical benefits of focusing on these graph classes. 
This improvement is particularly relevant for applications in data-intensive domains, where dense structures are common and efficiency gains are crucial.
William Lochet, Daniel Lokshtanov, Saket Saurabh, and Meirav Zehavi [STACS 2021]  harnessed the power of random samples as well as structure theory to obtain subexponential-time parameterized algorithms for {\sc Edge Odd Cycle Transversal}, {\sc Minimum Bisection}, and {\sc $d$-Way Cut}.
We construct a subexponential algorithm for {\sc UCED} by mapping the problem to {\sc $d$-Way Cut} and exploiting their algorithm. 

We also consider a relatively new graph editing operation called vertex splitting. 
Based on the two variants of vertex splitting operation, we study \textsc{Uniform Cluster Inclusive Vertex Splitting (UCIVS)} and 
\textsc{Uniform Cluster Exclusive Vertex Splitting (UCEVS)}.
A vertex split is a graph operation where a vertex \( v \) is replaced by two copies, with the combined neighborhood of the two copies being equal to the original neighborhood of \( v \). 
We provide a vertex kernel of size $4k$ and $4k$ for \textsc{Uniform Cluster Exclusive Vertex Splitting (UCEVS)} and 
\textsc{Uniform Cluster Inclusive Vertex Splitting (UCIVS)}, respectively.
\end{abstract}

\newpage

\section{Introduction}\label{Intro}

Graph modification problems, which aim to transform a graph into a desired structure through vertex or edge operations, have long been central to advances in theoretical computer science and parameterized complexity. 
Many well-known classical graph problems, such as \textsc{Vertex Cover}, \textsc{Feedback Vertex Set}, and \textsc{Odd Cycle Transversal}, can be framed as graph editing problems, where the objective is to modify a graph into a target graph class.
A prominent class of problems in this area is clustering, which has numerous applications in fields ranging from network design to data analysis.

In clustering, many real-world challenges involve grouping objects or data points into well-defined clusters.
By representing these challenges as graphs, where nodes represent objects and edges represent relationships, we can utilize graph modification techniques to enforce structural properties, such as equal-sized clusters or clusters containing highly similar elements. 
The ability to edit graphs—by deleting or adding edges, or by removing vertices—facilitates more accurate or computationally feasible clustering, particularly when the initial data does not naturally conform to ideal clusters.

In this paper, we study a relatively new variant of clustering problem.
We focus on computational problems where the goal is to modify a given graph through vertex deletions, vertex splittings, edge deletions, edge additions, or edge edits to transform it into a collection of equal-sized cliques, which we refer to as \textit{Uniform Cluster graphs}. 
Specifically, we investigate the {\sc Uniform Cluster Vertex Deletion (UCVD)} problem, where the task is to delete at most \( k \) vertices to achieve this transformation. 
Similarly, we consider the {\sc Uniform Cluster Vertex Splitting (UCVS)} problem, where the task is to decide whether there is a sequence of at most \( k \) vertex splits to get a uniform cluster graph.
Additionally, we consider the edge-based variants: {\sc Uniform Cluster Edge Deletion (UCED)}, {\sc Uniform Cluster Edge Addition (UCEA)}, and {\sc Uniform Cluster Edge Editing (UCEE)}, where the objective is to apply a bounded number of these operations to yield a graph that becomes a Uniform Cluster graph. We formally define the problems studied in this paper below.

\vspace{3mm}
\begin{center}
\fbox{\begin{minipage}{38.7em}\label{}
   \textsc{Uniform Cluster Vertex Deletion (UCVD)}\\
   \textbf{Input:} An undirected graph \( G=(V,E) \) and a positive integer \( k \).\\
   \textbf{Question:} Is there a subset \( S\subseteq V \) with \( |S|\leq k \) such that \( G-S \) is a collection of equal-sized cliques? 
\end{minipage} }
\end{center}

\begin{center}
\fbox{\begin{minipage}{38.7em}\label{}
   \textsc{Uniform Cluster Inclusive Vertex Splitting (UCIVS)}\\
   \textbf{Input:} An undirected graph \( G=(V,E) \) and a positive integer \( k \).\\
   \textbf{Question:} Is there a sequence of at most $k$ inclusive vertex splits that transforms $G$ into a collection of equal-sized cliques?
\end{minipage} }
\end{center}
\begin{center}
\fbox{\begin{minipage}{38.7em}\label{}
   \textsc{Uniform Cluster Exclusive Vertex Splitting (UCEVS)}\\
   \textbf{Input:} An undirected graph \( G=(V,E) \) and a positive integer \( k \).\\
   \textbf{Question:} Is there a sequence of at most $k$ exclusive vertex splits that transforms $G$ into a collection of equal-sized cliques?
\end{minipage} }
\end{center}
\begin{center}
\fbox{\begin{minipage}{38.7em}\label{EEEP}
   \textsc{Uniform Cluster Edge Editing (UCEE)}\\
   \textbf{Input:} An undirected graph \( G=(V,E) \) and a positive integer \( k \).\\
   \textbf{Question:} Is there a subset \( F\subseteq \binom{V(G)}{2} \) with \( |F|\leq k \) such that \( G\triangle F =(V,E\Delta F) \) is a collection of equal-sized cliques?     
\end{minipage} }
\end{center}
\begin{center}
\fbox{\begin{minipage}{38.7em}\label{EEDP}
   \textsc{Uniform Cluster Edge Deletion (UCED)}\\
   \textbf{Input:} An undirected graph \( G=(V,E) \) and a positive integer \( k \).\\
   \textbf{Question:} Is there a subset \( F\subseteq E \) with \( |F|\leq k \) such that \( G-F=(V,E\setminus F) \) is a collection of equal-sized cliques?     
\end{minipage} }
\end{center}
\begin{center}
\fbox{\begin{minipage}{38.7em}\label{EEEP1}
   \textsc{Uniform Cluster Edge Addition (UCEA)}\\
   \textbf{Input:} An undirected graph \( G=(V,E) \) and a positive integer \( k \).\\
   \textbf{Question:} Is there a subset \( F\subseteq \binom{V(G)}{2} \) with \( |F|\leq k \) such that \( G+F =(V,E \cup F) \) is a collection of equal-sized cliques?     
\end{minipage} }
\end{center}
\vspace{3mm}

\noindent{\bf Parameterized Complexity of Clustering problems in literature.} A classic problem in this domain is the \textsc{Cluster Vertex Deletion} problem, where the objective is to delete at most \( k \) vertices from a graph such that the remaining graph becomes a \textit{cluster graph}. A cluster graph is defined as a graph in which every connected component is a clique (i.e., a complete graph). The parameterized version of this problem has been the subject of extensive study, leading to the development of the best-known FPT algorithm with a running time of \( \mathcal{O}^{*}(1.811^k) \) \footnote{The \(\mathcal{O}^{*}(f(k))\) notation hides polynomial factors depending on \(n\), where \(n\) is the input size.}\cite{TsurD}, and a kernel of size \( \mathcal{O}(k^2) \) \cite{ABUKHZAM2010524}. Earlier contributions to this line of work have explored various facets of the problem, including complexity analysis and algorithmic strategies \cite{Boral2013AFB,CAI1996171,FERNAU2009177,Gramm2004AutomatedGO,10.1007/978-3-540-78773-0_61,Wahlstrom23420}.

Recently, Firbas et al. \cite{10.1007/978-3-031-52113-3_16} studied a problem called \textsc{Cluster Vertex Splitting} (CVS). The task in this problem is to transform an input graph into a cluster graph by performing at most \( k \) vertex splitting operations. A vertex split is a graph operation where a vertex \( v \) is replaced by two copies,say $v_1$ and $v_2$, with the combined neighborhood of the two copies being equal to the original neighborhood of \( v \). Abu-
Khzam et al. \cite{MR3835926} propose two different vertex splitting operations: one (exclusive
splitting) where $v_1$ and $v_2$ are required to have disjoint neighborhoods and another (inclusive splitting) where they are allowed to share neighbors. Intuitively, given a graph \( G \), the goal is to find cliques that cover all the edges of the graph with limited overlap between them. It has been shown that \textsc{CVS} admits a kernel with at most \( 3k+3 \) vertices.

Another notable problem in clustering is the \textsc{Cluster Editing} problem, also known as \textsc{Correlation Clustering}. In this problem, the goal is to decide whether a graph can be transformed into a cluster graph by adding or deleting at most \( k \) edges. The problem was first introduced by Ben-Dor, Shamir, and Yakhini \cite{doi:10.1089/106652799318274}, who were motivated by challenges in computational biology, and independently by Bansal, Blum, and Chawla, who applied the problem to document clustering in machine learning. Various parameterized versions of \textsc{Cluster Editing} and its many variants have since been intensively studied in the literature \cite{10.1007/978-3-642-25011-8_7,DBLP:conf/apbc/BockerBBT08,10.1007/978-3-540-68552-4_22,clusterfaster,BODLAENDER20101202,10.1007/s00224-008-9130-1,FELLOWS20112,10.1007/s00224-004-1178-y,DBLP:journals/algorithmica/GuoKKU11,doi:10.1137/090767285,10.1007/978-3-642-18381-2_29,10.1007/s00224-007-9032-7}. The problem is solvable in time \( \mathcal{O}^{*}(1.62^k) \), and it admits a kernel with at most \( 2k \) vertices.

\vspace{3mm}

\noindent{\bf Parameterized Complexity of Uniform Clustering problems in literature.} 
Misra et al. \cite{misra_et_al:LIPIcs.ISAAC.2023.53} examined this problem in a different context, investigating the complexity of reducing the number of distinct eigenvalues of an input graph by removing vertices or editing edges.
A special case where the number of eigenvalues is reduced to two is equivalent to transforming a graph into a Uniform Cluster graph, as shown in previous works \cite{Doob1970OnCC, GOLDBERG2020163}. It is proved that an adjacency matrix of a graph has at most two distinct eigenvalues if and only if the graph is a disjoint union of equal-sized cliques. 
Misra et al. provided various results for this special case.
Their study focused on four operations: vertex deletion, edge deletion, edge addition, and edge editing. 
The vertex deletion variant was shown to be NP-complete even on triangle-free and \( 3d \)-regular graphs for any \( d \geq 2 \), and NP-complete on \( d \)-regular graphs for any \( d \geq 8 \). 
Furthermore, the edge deletion, addition, and editing variants were also proven to be NP-complete.
The \textsc{UCEA} was shown to be NP-complete when the input is either a cluster graph, a forest, or a collection of cycles.
In addition to studying classical complexity, the authors provided numerous results in the realm of parameterized complexity, including FPT algorithms for the \textsc{UCVD} and \textsc{UCED} problems with running times of \( \mathcal{O}^{*}(3^{k}) \) and \( \mathcal{O}^{*}(2^{k}) \), respectively. These algorithms are based on the well known branching technique.

Furthermore, Madathil and Meeks recently explored a generalization of this problem termed \textsc{Balanced Cluster Editing}, where, given a graph \( G \), and two integers \( 0 \leq \eta \leq n \) and \( k \), we are permitted to edit up to \( k \) edges (see \cite{madathil2024parameterized}). 
This editing should yield a cluster graph where the size difference between any two connected components does not exceed \( \eta \). 
Notably, when \( \eta = 0 \), this problem reduces to the \textsc{Uniform Cluster Editing} problem.
Their paper presented polynomial kernels for \textsc{Balanced Cluster Completion}, \textsc{Balanced Cluster Deletion}, and \textsc{Balanced Cluster Editing}.
While they provided polynomial kernels for the general cases, we have achieved better bounds in all considered variants when \( \eta = 0 \). 
Our results were obtained simultaneously and independently from those by Madathil and Meeks.

\subsection{Our Results}

\noindent {\bf Parameterized Complexity of \textsc {Uniform Cluster Vertex Deletion}.}
Our contribution is a kernelization and an FPT algorithm.
Using simple yet powerful structural observations and non-trivial ideas, we develop combinatorial algorithms for the problem.
We prove that {\sc UCVD} admits a kernel with \( \mathcal{O}(k^{3}) \) vertices.
Recall that there is a $\mathcal{O}(k^{2})$ size vertex kernel for {\sc cluster vertex deletion}.
This is done by mapping the problem to {\sc $3$-Hitting set} problem.
As {\sc $d$-Hitting set} problem admits a kernel of size $\mathcal{O}(k^{d-1})$, we get the required kernel.
Note that in both {\sc Cluster Vertex Deletion} and {\sc Uniform Cluster Vertex Deletion}, we need to destroy all the induced $P_{3}$'s in a graph. 
If the graph is induced $P_{3}$-free then each connected component is a clique, that is, the graph is a cluster graph.
It is easy to see that {\sc Uniform Cluster Vertex Deletion} can be solved in polynomial time on cluster graphs by guessing the size of cliques in the resulting uniform cluster graph.
Although this is true, we observe that a reduction rule for {\sc Cluster Vertex Deletion} may not be applicable for 
{\sc Uniform Cluster Vertex Deletion} and vice a versa. 
For example, if a vertex is not in any induced $P_{3}$ then we can delete such vertex from the graph to obtain an equivalent instance of {\sc Cluster Vertex Deletion} but the same does not hold for {\sc Uniform Cluster Vertex Deletion}. 
Therefore we have to rely on newly constructed reduction rules for {\sc Uniform Cluster Vertex Deletion}.
We believe that the above observation is one of the important reason why we get a cubic kernel for {\sc Uniform Cluster Vertex Deletion} rather than a quadratic kernel like {\sc Cluster Vertex Deletion}. 

\par We present an FPT algorithm that runs in time \( \mathcal{O}^{*}(2^{k}) \), improving upon the FPT algorithm given in \cite{misra_et_al:LIPIcs.ISAAC.2023.53}, which has a running time of \( \mathcal{O}^{*}(3^{k}) \). 
We rely on \( \mathcal{O}^{*}(1.811^k) \) time algorithm for \textsc {Cluster Vertex Deletion} given in the literature.
If the algorithm returns a no instance then we can return a no instance for \textsc {Uniform Cluster Vertex Deletion}. 
If the algorithm returns a cluster vertex deletion set $X$ of size at most $k$ then we get that $G-X$ is a cluster graph. 
Here, we use a standard approach of guessing the intersection of solution with $X$ and then solving the annotated problem.

\vspace{3mm}

\noindent {\bf Parameterized Complexity of \textsc {Uniform Cluster Vertex Splitting}.}
In this work, we consider two kinds of vertex splitting operations called inclusive and exclusive vertex splitting.
An inclusive vertex split is a graph operation where a vertex \( v \) is replaced by two copies, with the combined neighborhood of the two copies being equal to the original neighborhood of \( v \). 
In the exclusive vertex splitting, the two copies of vertex must have no common neighbor.
We show that the \textsc {Uniform Cluster Inclusive Vertex Splitting} and \textsc {Uniform Cluster Exclusive Vertex Splitting} are NP-hard. 
In the realm of parametrized complexity, we provide a linear vertex kernel of size $4k$ for \textsc {Uniform Cluster Exclusive Vertex Splitting}
and a linear vertex kernel of size $4k$ for \textsc {Uniform Cluster Inclusive Vertex Splitting}.
Our NP-hardness as well as the kernelization result for {\sc UCEVS} relies on mapping the problem on another problem called {$K_{d}$-edge partition}.
In the {$K_{d}$-edge partition} problem, given a graph $G$, we need to determine if there is a partition of edges such that the graph induced by edges in every part is a clique of size $d$.
Our kernel also provides $k^{\mathcal{O}(k)} \cdot n^{\mathcal{O}(1)}$ time algorithm but it remains an open problem whether it can be improved to 
$2^{\mathcal{O}(k)} \cdot n^{\mathcal{O}(1)}$. 
Note that a similar question is also open for \textsc {Cluster Vertex Splitting}.
\vspace{3mm}

\noindent {\bf Parameterized Complexity of \textsc {Uniform Cluster Edge Deletion}, \textsc {Uniform Cluster Edge Addition} and \textsc {Uniform Cluster Edge Editing}.}
For the problem \textsc{UCEE}, we provide a kernel with \( \mathcal{O}(k^{2}) \) vertices. 
Furthermore, we present linear kernels with \( 6k \) and \( 5k \) vertices for \textsc{UCED} and \textsc{UCEA}, respectively, improving upon the quadratic kernel \cite{misra_et_al:LIPIcs.ISAAC.2023.53} for \textsc{UCEA}.
We also give an FPT algorithm with a running time of \( \mathcal{O}^{*}(1.47^{k}) \) for \textsc{UCED}, further enhancing the \( \mathcal{O}^{*}(2^{k}) \) algorithm \cite{misra_et_al:LIPIcs.ISAAC.2023.53}. 
This algorithm is based on the branching technique and uses branching rules given by B$\ddot{o}$cker and Damaschke \cite{clusterfaster} to construct FPT algorithm for {\sc Cluster Edge Deletion}.
One of the interesting as well as important class of graphs on which clustering problems are interesting are everywhere dense graphs. 
Everywhere-$\alpha$-dense graphs are graphs where each vertex has degree at least $\alpha n$ where $0 < \alpha \leq 1$ and $n$ denotes the number of vertices of graph.
Lochet et. al. \cite{lochet_et_al:LIPIcs.STACS.2021.50} provided subexponential algorithms for a number of classical problems such {\sc Minimum Bisection}, {\sc Edge Odd Cycle Transversal} and {\sc $d$-way cut}. All of these algorithms use a similar technique based on sampling.
Instead of going into details of this technique, we provide a reduction from \textsc {Uniform Cluster Edge Deletion} to {\sc $d$-way cut}. 
This allows us to construct a subexponential algorithm for \textsc {Uniform Cluster Edge Deletion} on everywhere-$\alpha$-dense graphs.\\

\par Through this work, we resolve three open questions posed by Misra et al. \cite{misra_et_al:LIPIcs.ISAAC.2023.53} concerning the complexity of these problems parameterized by solution size, thereby providing a clearer picture of the parameterized complexity landscape for the problems under consideration. 
The Table \ref{table:uniform_cluster_results} summarized the results given in this paper.

\vspace{3mm}

\begin{table}[ht]
\centering
\caption{FPT algorithms and kernel sizes for Uniform Cluster graph problems}
\label{table:uniform_cluster_results}
\begin{tabularx}{.9\textwidth} { 
  | >{\raggedright\arraybackslash}X 
  | >{\centering\arraybackslash}X 
  | >{\raggedleft\arraybackslash}X | }
   \hline
Uniform cluster  & FPT algorithm & Kernel \\
 \hline
Vertex Deletion & Theorem (\ref{FPTevd}) $\mathcal{O}^{*}(2^{k})$ & Theorem (\ref{UCVD}) $\mathcal{O}(k^{3})$ \\
\hline 
Exclusive Vertex Splitting & - & Theorem (\ref{UCEVSkernel}) $4k$ \\
\hline 
Inclusive Vertex Splitting & - & Theorem (\ref{UCIVSkernel}) $4k$ \\
 \hline
 Edge Editing  & - &   Theorem (\ref{UCEE}) $\mathcal{O}(k^{2})$  \\
\hline
Edge Deletion  &  Theorem (\ref{FPT-UCED}) $\mathcal{O}^{*}(1.47^{k})$ &  Theorem (\ref{kernel-UCED})  $6k$  \\
\hline
Edge Addition  & - &   Theorem (\ref{kernel-UCEA}) $5k$  \\
\hline
\end{tabularx}
\end{table}

\vspace{5mm}




\subsection{Organization of the Paper}

In \textbf{Section 1}, we provide an introduction, a survey of related work, an outline of our results, and a high-level technical overview.  
\textbf{Section 2} focuses on the parameterized complexity of the \textsc{Uniform Cluster Vertex Deletion} problem.  
In \textbf{Section 3}, we present our results on the parameterized complexity of two vertex-splitting variants: \textsc{Uniform Cluster Inclusive Vertex Splitting} and \textsc{Uniform Cluster Exclusive Vertex Splitting}.  
\textbf{Section 4} discusses results on the parameterized complexity of edge-based variants.  
Finally, \textbf{Section 5} concludes the paper, summarizing the key findings and identifying potential directions for future research.

\subsection{Technical Overview of Proof of Theorem \ref{UCVD}}\label{sec:tech-overview}

The proof of Theorem \ref{UCVD} is intricate and relies on a combination of techniques and ideas, making it considerably more challenging compared to the proofs of other results presented in this paper. To assist the reader in navigating this complexity, we provide an overview of the proof in this section.

This overview outlines the key steps, insights, and high-level strategies used in the proof without delving into the technical details, which are reserved for the formal presentation. The goal is to give the reader a roadmap of the argument and highlight the logical flow.

In contrast, the proofs of other results in this paper are relatively straightforward and can be understood directly from the formal statements and derivations. For these results, no additional overview is necessary.

\begin{theorem}\label{UCVD}
   The {\sc UCVD} problem parameterized by solution size admits a kernel of size $\mathcal{O}(k^3)$.
\end{theorem}

\noindent To prove this theorem, we begin by greedily constructing a maximal set $\mathcal{P}_{3}$ of vertex disjoint induced $P_{3}$s in $G$.
Let $S$ be the vertices of this set. 
Clearly, $|S|\leq 3k$ as otherwise we can return a no instance.
This implies that $G-S$ is a cluster graph. 
We denote the set of cliques of $G-S$ by  $\mathcal{C}$.

We begin with a reduction rule that helps us bound the number of cliques in $\mathcal{C}$: if a vertex $s \in S$ has neighbors in at least $k+2$ distinct cliques in $G - S$ or 
$s$ has at least $k+1$ neighbors in a clique $C \in \mathcal{C}$ and more than $k$ neighbors in cliques other than $C$ then $s$ must be part of the solution. 
This is because in this situation, it is not possible to destroy all induced $P_{3}$s by deleting at most $k$ vertices without deleting $s$. 
This allows us to bound the number of cliques in $G - S$ by $\mathcal{O}(k^2)$.

Next, we consider the size of the largest clique in $G - S$. If this size is bounded by $8k$, we obtain the desired kernel immediately.
In the other case, we assume that the size of a largest clique has a lower bound by $8k$ for some sufficiently large fixed constant. 
A key observation here is that the sizes of cliques in $G - S$ can differ by at most $4k$.
This follows from the fact that we are allowed to delete at most $k$ vertices, and since $|S| \leq 3k$, this limits the number of vertices that can be edited within the cliques.
As a result, all cliques in $G - S$ are bounded within a fixed size interval of $4k$.
Given that all cliques in $G - S$ are sufficiently large (with size at least $4k$), each vertex $s \in S$ must either be part of the solution or belong to an unique clique in $G - S$ after deleting the vertices in a solution. 
This leads us to define the notion of a heavy neighbor for each vertex $s \in S$. 
A clique $C$ in $G - S$ is called a heavy neighbor of $s$ if $s$ is adjacent to at least $\max\{|C| - 4k, k+1\}$ vertices in $C$. 
The (first) reduction rule given earlier shows that each vertex $s \in S$ has exactly one heavy neighbor. 
Consequently, $s$ can have at most $k$ neighbors in cliques other than its heavy neighbor.

Since the largest clique in $G - S$ is at least $8k$ in size, the equal-sized cliques obtained after deleting a solution must have size at least $7k$.
This implies that no subset of vertices from $S$ can independently form a clique in the final graph without incorporating vertices from $G - S$.
Consequently, after deleting the solution, every remaining vertex in $S$ must join a unique clique (its heavy neighbor) in $G - S$.

Next, we provide one of the most crucial reduction rule in this proof which helps us bound the size of each clique in $\mathcal{C}$ by $\mathcal{O}(k^{2})$.
This rule is based on the following key insight: if every clique $C \in \mathcal{C}$ contains at least $k+2$ true twins, we can safely delete one vertex from each such set in every clique.
The correctness of this idea follows from the following observation.
Note that assuming we are dealing with yes instance, these deleted vertices can be added back to the final uniform cluster graph without violating the equal-sized clique structure. 
This holds because the remaining true twin ensures that the clique remains intact when a vertex is reinserted. 
Furthermore, since the same number of vertices are deleted from each clique, reinserting them preserves the equal sizes of the cliques.
This reduction rule, combined with previous observations, allows us to effectively bound the size of the cliques in $\mathcal{C}$ by $\mathcal{O}(k^2)$.
This bounds the total number of vertices in the graph by $\mathcal{O}(k^{4})$.
Note that there are at most $3k$ cliques in $\mathcal{C}$ which are heavy neighbours of vertices in $S$.
Finally, a series of simple reduction rules helps us bound the the cliques of large size (containing at least $\mathcal{O}(k^{2})$ vertices) which are not heavy neighbour of any vertex in $S$  by $2k+1$.
This allows us to get the desired kernel.

\subsection{Preliminaries}
Throughout this article, $G=(V,E)$ denotes a finite, simple and undirected graph of order $|V|=n$. For  $u\in V$, we define $N(u)=\{v\in V : (u,v) \in E\}$ and $N[u]=N(u) \cup \{u\}$. The  \emph{degree} of $u \in V$ is $|N(u)|$ and denoted by $d_G(u)$.  A  \emph{clique} $C$ in an undirected graph  $G = (V, E)$ is a subset of the vertices $C \subseteq V$ such that every two distinct vertices are adjacent. 
A \emph{cluster graph} is a
graph where every component is a clique. Observe that a graph is a cluster graph
if and only if it does not have an induced $P_3$, that is, an induced path on three vertices.  Let $U\subseteq V$ be a subset of 
vertices of $G$ and $F\subseteq \binom{V}{2}$ be a subset of pairs of vertices of $G$.
The subgraph induced by $U \subseteq V$ is denoted by $G[U]$.
We define $G-U= G[V\setminus U]$, $G-F=(V,E\setminus F)$, $G+F=(V,E\cup F)$ and 
$G\triangle F=(V,E\triangle F)$. Here, $E\triangle F$ is the symmetric difference between $E$ and $F$. If $U=\{u\}$ or $F=\{e\}$ then we simply write $G-u$, $G-e$ and $G+e$ for 
$G-U$, $G-F$ and $G+F$, respectively. Next, we discuss the editing operation called vertex splitting.

\begin{definition}[Inclusive Vertex Splitting]
Let \( G = (V, E) \) be a graph and \( v \in V \) a vertex. An \textit{inclusive vertex split} replaces \( v \) with two new vertices \( v_1 \) and \( v_2 \), such that:
\begin{itemize}
    \item \( N(v_1) \cup N(v_2) = N(v) \), where \( N(v) \) is the neighborhood of \( v \);
    \item \( N(v_1) \cap N(v_2) \) may be non-empty, allowing \( v_1 \) and \( v_2 \) to share neighbors.
\end{itemize}
\end{definition}

\begin{definition}[Exclusive Vertex Splitting]
An \textit{exclusive vertex split} also replaces \( v \) with two new vertices \( v_1 \) and \( v_2 \), but with stricter conditions:
\begin{itemize}
    \item \( N(v_1) \cup N(v_2) = N(v) \);
    \item \( N(v_1) \cap N(v_2) = \emptyset \), ensuring that \( v_1 \) and \( v_2 \) have disjoint neighborhoods.
\end{itemize}
\end{definition}

\begin{figure}[ht]
    \centering
    \begin{subfigure}[b]{0.3\textwidth}
        \centering
        \begin{tikzpicture}[scale=1, every node/.style={circle, draw, fill=white, inner sep=2pt}]
            \node (a) at (0, 1) {a};
            \node (b) at (-1, 0) {b};
            \node (c) at (1, 0) {c};
            \node (d) at (0, -1) {d};
            
            \draw (a) -- (b);
            \draw (a) -- (c);
            \draw (a) -- (d);
            \draw (b) -- (c);
        \end{tikzpicture}
        \caption{Original Graph}
        \label{fig:original}
    \end{subfigure}
    \hfill
    \begin{subfigure}[b]{0.3\textwidth}
        \centering
        \begin{tikzpicture}[scale=1, every node/.style={circle, draw, fill=white, inner sep=2pt}]
            \node (a1) at (-0.5, 1) {a$_1$};
            \node (a2) at (0.5, 1) {a$_2$};
            \node (b) at (-1, 0) {b};
            \node (c) at (1, 0) {c};
            \node (d) at (0, -1) {d};
            
            \draw (a1) -- (b);
            \draw (a1) -- (c);
            \draw (a2) -- (c);
            \draw (a2) -- (d);
            \draw (b) -- (c);
        \end{tikzpicture}
        \caption{Inclusive Vertex Splitting}
        \label{fig:inclusive}
    \end{subfigure}
    \hfill
    \begin{subfigure}[b]{0.3\textwidth}
        \centering
        \begin{tikzpicture}[scale=1, every node/.style={circle, draw, fill=white, inner sep=2pt}]
            \node (a1) at (-0.5, 1) {a$_1$};
            \node (a2) at (0.5, 1) {a$_2$};
            \node (b) at (-1, 0) {b};
            \node (c) at (1, 0) {c};
            \node (d) at (0, -1) {d};
            
            \draw (a1) -- (b);
            \draw (a1) -- (c);
            \draw (a2) -- (d);
            \draw (b) -- (c);
        \end{tikzpicture}
        \caption{Exclusive Vertex Splitting}
        \label{fig:exclusive}
    \end{subfigure}
    
    \caption{Illustration of Inclusive and Exclusive Vertex Splitting. (a) The original graph \( G \) with vertex \( a \) connected to \( b \), \( c \), and \( d \). (b) Inclusive splitting allows \( a_1 \) and \( a_2 \) to share neighbors and optionally connect to each other. (c) Exclusive splitting divides the neighbors of \( a \) between \( a_1 \) and \( a_2 \) with disjoint neighborhoods.}
    \label{fig:vertex_splitting}
\end{figure}

\noindent The key difference between these two modifications lies in their treatment of vertex neighborhoods. In Inclusive Vertex Splitting, $v_1$ and $v_2$ are allowed to share neighborhoods, while in Exclusive Vertex Splitting, $v_1$ and $v_2$ must have disjoint neighborhoods.

For the above two definitions, we say $v$ was split into $v_1$ and $v_2$, and call these vertices the descendants of $v$. Conversely, $v$ is called the ancestor of $v_1$ and $v_2$.

\begin{definition}
    A \textit{splitting sequence} of $k$ splits is a sequence of graphs $G_0, G_1, \ldots, G_k$, such that $G_{i+1}$ is obtainable from $G_i$ via a vertex split for $i \in \{0, \dots, k-1\}$.
\end{definition}
The notion of \textit{descendant vertices} (resp. \textit{ancestor vertices}) is extended in a 
transitive and reflexive way to splitting sequences.

\vspace{3mm}

In the example given below, we demonstrate that inclusive vertex splitting is a stronger operation than exclusive vertex splitting. In the following graph, we can get a disjoint union of triangles by doing two inclusive vertex splitting operations.
But if we allow only exclusive vertex splitting, then we need at least $4$ vertex splitting operations to get a uniform cluster graph from $G$.    

\begin{center}

\begin{tikzpicture}

\centering
	\node[circle,draw, minimum size=0.1cm ] (a) at (-3, 2) [label=above:$a$]{};
	\node[circle,draw, minimum size=0.1cm] (b) at (-1.5,2)  [label=above:$b$]{};
	\node[circle,draw, minimum size=0.1cm] (c) at (-3, 0.5) [label=below:$c$]{};
	\node[circle,draw, minimum size=0.1cm] (d) at (-1.5, 0.5) [label=below:$d$]{};
	
	\node[] (a2) at (-2.2, 0) [label=below:${G}$]{};

\path 
	
(a) edge (b)
(a) edge (c)	
(c) edge (d)	
(b) edge (d)	
(b) edge (c);
 \end{tikzpicture}
 \end{center}

\section{Kernelization and FPT algorithm for {\sc UCVD} parameterized by solution size}\label{section UCVD}

In this section, we provide kernelization and fixed-parameter tractable (FPT) algorithm for the Uniform Cluster Vertex Deletion (UCVD) problem.

\subsection{Proof of Theorem \ref{UCVD}}
     
\noindent A cluster graph is a graph in which every connected component is a clique. 
Note that a graph is a cluster graph if and only if it does not contain an induced 
$P_3$. 
The first step of the kernelization process is to compute a maximal set $\mathcal{P}_3$
of vertex-disjoint induced $P_3$s in $G$.  
If $|\mathcal{P}_3|>k$, we have a no-instance.
So we assume that $|\mathcal{P}_3|\leq k$, and let $S$ be the vertices of $\mathcal{P}_3$.
We have $|S|\leq 3k$. Let us denote the set of cliques of $G-S$ by 
$\mathcal{C}$.\\

\noindent We begin with a following reduction rule that provides two sufficient conditions for a vertex $s\in S$ to be in a solution.

\begin{evd}\label{Red1} Let $s \in S$ be a vertex such that one of the following conditions hold: 
\begin{itemize} \item $s$ has neighbors in at least $k+2$ distinct cliques in $\mathcal{C}$, or 
\item $s$ has at least $k+1$ neighbors in a clique $C \in \mathcal{C}$ and more than $k$ neighbors in cliques other than $C$. 
\end{itemize} 
Then, delete $s$ from $G$ and decrement the parameter $k$ by 1. The new instance becomes $(G - s, k - 1)$. 
\end{evd}

\begin{lemma}
  Reduction Rule UCVD \ref{Red1} is safe.   
\end{lemma}

\proof Let $X$ be a uniform cluster vertex deletion set of $G$ of size at most $k$. 
For the sake of contradiction, assume that $s \notin X$.
In all cases outlined by Reduction Rule \ref{Red1}, the graph $G - X$ will contain a $P_3$ containing $s$, violating the structure of a (uniform) cluster graph. 
This contradicts the assumption that $X$ is a valid uniform cluster vertex deletion set. 
Hence, $s$ must belong to any solution for $(G, k)$, and it is safe to delete $s$ and decrement $k$ by 1. \qed

\vspace{3mm}

\noindent Next, we partition the set of cliques in $\mathcal{C}$ based on  whether they have neighbours in $S$ or not. We define: 

$$\mathcal{C}_0= \{C \in  \mathcal{C} :  \text{no vertex in } C \text{ has  a neighbour in } S\}$$  and 
$$\mathcal{C}_{1}= \{C \in \mathcal{C}: \text{some vertex
in } C  \text{ has  a  neighbour in }  S \}.$$

\noindent An $r$-clique is a clique of size $r$. 
Now, we have the following simple rule.

\begin{evd}\label{Red2} If  for some $r\in \{1,2,\ldots,n\}$ there are more than $k+1$  
$r$-cliques in $\mathcal{C}_0$, then remove all but $k+1$ $r$-cliques from $G$. 
\end{evd}

\begin{lemma}
Reduction rule UCVD \ref{Red2} is safe. 
\end{lemma}

\proof Suppose $(G,k)$ is a yes-instance  and 
$X$ is a uniform cluster vertex deletion set of $G$ of size at most $k$.
If  for some $r\in \{1,2,\ldots,n\}$ there are more than $k+1$  
$r$-cliques in $\mathcal{C}_0$, then we claim that $r$ is 
the size of all cliques in $G-X$. For the sake of contradiction, let us assume that $r'<r$ is  the size of all cliques in $G-X$. In that case, we  need to delete at least one vertex  from each  $r$-clique.  This means we would have to add at least  $k+1$ vertices in $X$, which contradicts the assumption  that $X$ is of size at most $k$. 
Thus, we conclude that $r$ is the size of all cliques in $G-X$. Since the solution size is at most $k$, this allows us to remove all but $k+1$  $r$-cliques from $G$. \qed

\begin{evd}\label{Red3} If $\mathcal{C}_0$  contains cliques of more than $k+1$ distinct sizes then return a no instance.
\end{evd}

\noindent This is correct because by deleting at most $k$ vertices 
we can alter the sizes of at most $k$ cliques. 
Therefore, in this case, there can not exist a uniform cluster vertex deletion of size at most $k$.

\begin{lemma}
    After applying Reduction Rule \ref{Red1}, \ref{Red2} and \ref{Red3}, the set $\mathcal{C} = \mathcal{C}_{0} \cup \mathcal{C}_{1} $ contains at most $\mathcal{O}(k^{2})$ cliques.
\end{lemma}

\proof Due to Reduction Rule \ref{Red2} and \ref{Red3}, we know that $|\mathcal{C}_{0}|\leq (k+1)(k+1)$.
After the exhaustive application of reduction rule \ref{Red1},
each vertex of $S$ can have neighbours in at most $k+1$ cliques of $\mathcal{C}_{1}$. 
Since $|S|\leq 3k$ and each vertex of $S$ can have neighbours 
in at most $k+1$ cliques of $\mathcal{C}_{1}$,
there are at most $3k(k+1)$ cliques in $\mathcal{C}_{1}$.
Therefore, we have  $|\mathcal{C}_{1}|\leq 3k(k+1)$.
Combining the two, we get  $$|\mathcal{C}|=|\mathcal{C}_0|+|\mathcal{C}_{1}| \leq (k+1)(k+1)+3k(k+1)= 4k^{2}+5k+1.$$ \qed

\noindent Now, we consider two cases based on the maximum size of a clique in $\mathcal{C}$. Let $\omega(\mathcal{C}) = \max \limits_{C \in \mathcal{C}} \{|C|\}$. 

\paragraph{Case 1: $\omega(\mathcal{C}) < 8k$} In this case, we obtain a kernel of size:
\[
|S| + \sum\limits_{C \in \mathcal{C}} |C| < 3k + 8k(4k^2 + 5k + 1) = \mathcal{O}(k^3).
\]

\paragraph{Case 2: $\omega(\mathcal{C}) \geq 8k$} Let $C_0$ be a clique in $\mathcal{C}$ such that $\omega(\mathcal{C}) = |C_0|$. Then $C_0$ becomes a clique of size at least $|C_0| - k \geq 7k$ in $G - X$, where $X$ is a solution of size at most $k$. In this case, the resulting graph $G - X$ must be a collection of $r$-cliques, where $r \geq |C_0| - k \geq 7k$.

\begin{evd}\label{Red3a}
    If there exist $C_{1}$ and $C_{2} \in \mathcal{C}$  such that $ |C_1| -|C_{2}|  > 4k$, then delete all the vertices in $C_{2}$ and decrement the parameter $k$ by $|C_{2}|$. The new instance is $(G-C_{2},k-|C_{2}|)$.
\end{evd}

\begin{lemma}
    Reduction Rule UCVD \ref{Red3a} is safe,
\end{lemma}
\proof For the sake of contradiction, let us assume that  $v \in C_{2}$  does not belong to 
some solution $X$ of size at most $k$. 
First, since $C_{2}$ is a clique, $v$ has at most $|C_{2}|-1$ neighbours in $C_{2}$ in the resulting graph.
Second, $v$ can have at most $3k$ neighbours in $S$.
Thus, $v$ can be part of a clique of size at most   $(|C_{2}|-1)+3k+1= |C_{2}|+3k$.\\

On the other hand, there always exist a vertex $u\in C_{1}$ such that $u$ is not contained in $X$.
Clearly, $u$ will have at least $|C_{1}|-k-1$ neighbours in $G-X$.
Therefore, $u$ must be contained in a clique of size at least $|C_{1}|-k >
|C_{2}|+4k-k= |C_{2}|+3k$. 
Therefore, $u$ and $v$ can never be contained in equal-sized cliques.  This contradicts the fact that  $G-X$ is  a collection of equal sized-cliques. Therefore, we must include all vertices of $C_{2}$ in every uniform cluster 
vertex deletion set $X$ of $G$ of size at most $k$. \qed

\begin{observation}
    After applying Reduction Rule UCVD \ref{Red3a}, we get that $\min\limits_{C\in \mathcal{C}} |C| \geq 4k$.
\end{observation}
This is true because we have assumed that $\omega(\mathcal{C})\geq 8k$.

\vspace{5pt}
 \noindent We now introduce the concept of a \emph{heavy neighbour} of a vertex $s\in S$ in $\mathcal{C}$, which is a key element for explaining the forthcoming reduction rules.

\begin{definition}
 A clique $C\in \mathcal{C}$ is called a heavy neighbour of $s\in S$ if 
 $s$ is adjacent to at least $\max\{|C|-4k,k+1\}$ vertices in $C$.
\end{definition}

\noindent The next reduction rule helps to show that each vertex $s$ has a unique heavy neighbour in $\mathcal{C}$.

\begin{evd}\label{Red4}
    If $s\in S$ has no heavy neighbour, remove $s$ from $G$, and  decrement the parameter 
    $k$ by 1. The new instance is $(G-s,k-1)$.
\end{evd}

\begin{lemma}
Reduction rule UCVD \ref{Red4} is safe.
\end{lemma}
\proof For the sake of contradiction, let us assume that $s\in S$ has no heavy neighbour and  does not belong to some 
solution $X$ of size at most $k$. 
It means $s$ must belong to some clique  in $G-X$. Since $s$ has no heavy neighbour
in $\mathcal{C}$, $s$ is adjacent to at most $\max\{|C|-4k,k+1\}-1$ vertices in some $C\in \mathcal{C}$.
Additionally, $s$ can have at most $3k-1$ neighbours in $S$.  Now consider two cases:
\begin{enumerate}
    \item If $\max\{|C|-4k,k+1\}-1=|C|-4k-1$, then $s$ can be part of a clique of size at most   $(|C|-4k-1)+(3k-1)+1= |C|-k-1$. This contradicts the fact that  $G-X$
is  a collection of $r$-cliques, where $r\geq |C|-k$. 
\item If $\max\{|C|-4k,k+1\}-1=k$, then $s$ can be part of a clique of size at most   $ k+(3k-1)+1= 4k$. This contradicts the fact that  $G-X$
is  a collection of $r$-cliques, where $r\geq 7k$. 
\end{enumerate}
Therefore, $s$ must be part of every solution of size at most  $k$. \qed

\begin{lemma}
    After applying the Reduction rules \ref{Red1} and \ref{Red4} exhaustively, each $s$ has exactly one heavy neighbour.
\end{lemma}
\proof The condition 2 in the Reduction Rule \ref{Red1} makes sure that $s$ has at most one heavy neighbour. On the other hand, the Reduction rule \ref{Red4} makes sure that $s$ has at least one heavy neighbour. Therefore, it shows that each $s$ has exactly one neighbour. \qed

\noindent Let $S_{C}\subseteq S$ denotes the set of vertices in $S$ whose heavy neighbour is $C$.
For each $C \in \mathcal{C}$, we define

$$ \mathcal{N}(C) := \bigcup_{s\in S_{C}}\Big(C\setminus N(s)\Big) \cup \bigcup_{  s \in S\setminus S_{C}} 
{\Big(N(s)\cap C \Big)} \ \  \text{and} \ \  \mathcal{N}(\mathcal{C}) = \bigcup\limits_{C \in \mathcal{C}} \mathcal{N}(C).$$

\noindent 
Next, we show that the vertices of 
$ C\setminus \mathcal{N}(C)$ are true twins.

\begin{lemma}  
 For each $C\in \mathcal{C}$, every pair of adjacent vertices $x$ and $y$ in $C\setminus \mathcal{N}(C)$ are true twins, that is, $N[x]=N[y]$. 
\end{lemma}

\proof By the definition of $\mathcal{N}(C)$, note that for any vertex $v \in C\setminus \mathcal{N}(C)$, we have $N(v)\cap S= S_C$. Recall that for the 
vertices in $S_C$, $C$ is their  heavy neighbour. Since $v\in C$ and $C$ is a clique, $v$ is adjacent to all other vertices of $C$. Therefore, for every $v \in C\setminus \mathcal{N}(C)$, we have $ N[v] = C\cup S_C.$  \qed

\begin{evd}\label{Red6}
    If  $\min \limits_{C \in \mathcal{C}} \{|C\setminus \mathcal{N}(C)|\} > k+1$, then arbitrarily delete exactly one vertex from each $ C\setminus \mathcal{N}(C)$ for every $C\in \mathcal{C}$. 
    The new instance is $(G-W,k)$, where $W$ is the set of vertices deleted from  $G$. 
\end{evd}

\begin{lemma}
    Reduction Rule UCVD \ref{Red6} is safe.
\end{lemma}
\proof Let $I'=(G',k)$ be an instance of  {\sc UCVD} obtained from $I=(G,k)$
by exhaustively applying  Reduction Rule \ref{Red6}. We will show that $I$ is a yes-instance if and only if $I'$ is a 
yes-instance.

Let $X\subseteq V(G)$ be a solution  of size at most $k$ for the instance $I$. 
We will construct a solution $X'$ of the same size as $X$ for the instance $I'$. 
We know $G-X$ is a collection of equal sized cliques $Q_{1},Q_{2},\ldots,Q_{r}$. 
We observe that there is no clique  $Q_{i}$ such that $Q_{i}\subseteq S$. 
This is true because otherwise $|Q_{i}|\leq 3k$ as $|S|\leq 3k$, 
and we have already observed that the size of cliques must be at least $7k$. 
We construct  $X'$ from $X$ as follows: if $u\in X\cap(C\setminus \mathcal{N}(C))$ for some $C\in \mathcal{C}$ and 
$u$ is deleted during the execution of Reduction Rule \ref{Red6}, then remove 
$u$ from $X$ and include $u$'s true twin $v \notin X$. 
We have to show that $G'-X'$ is a collection of equal-sized cliques. It is easy to see that $G'-X'$ is a collection of cliques $Q'_i=Q_i\setminus \{u\}$ where $u \in Q_{i}$ is a vertex deleted by Reduction Rule \ref{Red6}. 
As $|Q_{i}|=|Q_{j}|$, we  get $|Q'_i|=|Q'_j|$ for all $i\neq j$. 
This shows that $I'$ is a yes-instance. 

In the other direction, let us assume that $X'$ is a solution of size at most $k$  for the instance $I'$. Let us denote the  equal sized cliques in $G'-X'$ by $Q'_{1},Q'_{2},\ldots,Q'_{r}$.
We claim that $X:=X'$ is a solution for the instance $I$.
Note that $G-X$ is again a collection of cliques $Q_{1},Q_{2},\ldots,Q_{r}$ where $Q_{i}=Q'_{i}\cup \{u\}$ where $u$ is a vertex deleted by Reduction Rule \ref{Red6}. 
Clearly, we have $|Q_{i}|=|Q_{j}|$  as we know that  $|Q'_{i}|=|Q'_{j}|$ for all $i\neq j$.\qed \\

\begin{lemma}
    After the exhaustive application of the above reduction rules, we have $\max\limits_{C\in \mathcal{C}} |C| \leq \mathcal{O}(k^{2})$.
\end{lemma}

\proof We begin by providing an upper bound on $|\mathcal{N}(\mathcal{C})|$.

\begin{claim}
    We have $|\mathcal{N}(\mathcal{C})|= \mathcal{O}(k^2)$.
\end{claim}
\proof Let us assume that $C_{s}$ denotes the unique heavy neighbour of $s$. We have 
\begin{align*}
\mathcal{N}(\mathcal{C}) & = \bigcup_{C\in \mathcal{C}} \Bigg[\bigcup_{s\in S_{C}}\Big(C\setminus N(s)\Big) \cup \bigcup_{  s \in S\setminus S_{C}} 
{\Big(N(s)\cap C \Big)}\Bigg]\\
 &= \bigcup_{s \in S} \Bigg[ \Big(C_{s}\setminus N(s)\Big) 
  \cup \bigcup_{C\in \mathcal{C}\setminus C_{s} } 
{\Big(N(s)\cap C \Big)}\Bigg]
\end{align*}
Note that by Reduction Rule \ref{Red1}, for any fixed $s\in S$, we have 
$$\bigg| \bigcup_{C\in \mathcal{C}\setminus C_{s} } 
{\Big(N(s)\cap C \Big)} \bigg|< (k+1).$$
Also, for each $s\in S$, we have 
\begin{align*}
\big|C_{s}\setminus N(s)\big|  &\leq |C|-(|C|-4k)=  4k \mbox{ if } \max\{|C|-4k,k+1\}=|C|-4k\\
 &\leq |C|-(k+1) \leq 4k \mbox{ if } \max\{|C|-4k,k+1\}=k+1
 \end{align*}
 
\noindent As there are at most $3k$ vertices in $S$, we get that 
$|\mathcal{N}(\mathcal{C})|\leq 3k[k+4k]=15k^{2}=\mathcal{O}(k^2).$ \claimqed

\vspace{3mm}

\noindent  After exhaustively applying  Reduction Rule \ref{Red6}, 
we know that there exists a clique, say $C_{q} \in  \mathcal{C}$, such that
$|C_q|-|\mathcal{N}(C_{q})| \leq k+1$. 
Note that $\max \limits_{C \in \mathcal{C}} |C|-|C_{q}| \leq 4k$ due to Reduction Rule UCVD \ref{Red3a}. 
Therefore, we get $\max\limits_{C\in \mathcal{C}} |C|- |\mathcal{N}(C_{q})|\leq 5k+1$. Since  $|\mathcal{N}(C_{q})|= \mathcal{O}(k^{2})$, we get $\max\limits_{C\in \mathcal{C}} |C| \leq \mathcal{O}(k^{2})$. \qed

\noindent Note that 

\[
V(G) = S \cup \bigcup_{C \in \mathcal{C}} C = S \cup \bigcup_{C \in \mathcal{C}} \big(\mathcal{N}(C) \cup (C \setminus \mathcal{N}(C))\big) = S \cup \mathcal{N}(\mathcal{C}) \cup \bigcup_{C \in \mathcal{C}} (C \setminus \mathcal{N}(C))
\]

\noindent To get a cubic kernel, it is enough to show that the size of the set $\bigcup\limits_{C \in \mathcal{C}} (C \setminus \mathcal{N}(C))$ is bounded by a cubic function of the parameter. 
To see this, we partition the set of cliques in  $\mathcal{C}$ into two parts:
$\mathcal{H}=\{ C\in \mathcal{C} ~|~ C \text{ is a heavy neighbour of some } s\in S \}$ and $\mathcal{L}=\mathcal{C}\setminus \mathcal{H}$. 
As we have seen, every vertex  $s\in S$ has exactly one heavy neighbour, which 
implies that $|\mathcal{H}|\leq 3k$. Therefore, we get that $\bigcup\limits_{C \in \mathcal{H}} |(C \setminus \mathcal{N}(C))|=\mathcal{O}(k^{3})$ as  $\max\limits_{C\in \mathcal{C}} |C| \leq \mathcal{O}(k^{2})$.
Now, let us focus on the set $\mathcal{L}$. Let us denote 
$\mathcal{L}^{\prime}=\{C \in \mathcal{L} ~|~ |C \setminus \mathcal{N}(C)|\geq k+1\}$ and $\mathcal{L}^{\prime \prime} = \mathcal{L} \setminus \mathcal{L}^{\prime}$. 
It is clear that  $\bigcup\limits_{C \in \mathcal{L}^{\prime \prime}} |(C \setminus \mathcal{N}(C))| \leq \mathcal{O}(k^{3})$ as $|\mathcal{L}^{\prime \prime}|\leq |\mathcal{C}| \leq \mathcal{O}(k^{2})$.
We can assume that $|\mathcal{L}^{\prime}| \geq 2k+1$; otherwise, we achieve the required cubic kernel.

\begin{lemma}
    If $|\mathcal{L}^{\prime}| \geq 2k+1$ then in polynomial time we can determine the size $c$ of equal sized cliques obtained after deleting the vertices in a solution, assuming the input is a 
    yes-instance.
\end{lemma}
\proof Let us denote $\mathcal{L}^{\prime} = \{C^{\prime}_{1}, C^{\prime}_{2},\ldots, C^{\prime}_{l}\}$, where $l \geq 2k+1$.  Let $X$ be a solution of size at most $k$ of the input instance. 
As $|C^{\prime}_{i} \setminus \mathcal{N}(C^{\prime}_{i})|\geq k+1$, at least one vertex will survive from each set $C^{\prime}_{i} \setminus \mathcal{N}(C^{\prime}_{i})$ in $G-X$.  Because the vertices in $C^{\prime}_{i} \setminus \mathcal{N}(C^{\prime}_{i})$ 
do not have any neighbours in $S$, the deletion of at most $k$ vertices can  change the degree of vertices in at most $k$ sets  $C^{\prime}_{i} \setminus \mathcal{N}(C^{\prime}_{i})$. 
This implies that  vertices in all but at most $k$ distinct  sets $C^{\prime}_{i} \setminus \mathcal{N}(C^{\prime}_{i})$ must have same degree. 
If this is not the case, we can return a no-instance. 
If it is the case, then let us say the degree is $d$.
One can easily see that $c=d+1$.
It is important to note that having $|\mathcal{L}^{\prime}| \geq 2k+1$ is necessary; otherwise, identifying $d$ is not possible. \qed

\noindent The following set of reduction rules are only applicable when $|\mathcal{L}^{\prime}| \geq 2k+1$.

\begin{evd}\label{Red7}
If there is a vertex $v$ such that $d(v)<d$, then add it to the solution. The new instance is $(G-v,k-1)$.
\end{evd}

\noindent Note that the degree of every vertex in $G-X$ must be equal to $d$. Since 
$d(v)<d$ and  we cannot increase the degree of $v$ through vertex deletion, $v$ must be added to the solution.
Due to Reduction Rule \ref{Red7}, we can assume that all the cliques in $\mathcal{L}^{\prime}$ have size at least $c$.

\begin{evd}\label{Red8}
If there is a clique $C'\in \mathcal{L}^{\prime}$ such that $|C^{\prime}|=c$, then delete all the vertices in $N(C')$ from the graph $G$. The new instance is $(G-N(C'),k-|N(C')|)$. 
\end{evd}

\begin{lemma}
    Reduction Rule \ref{Red8} is safe.
\end{lemma}
\proof Note that at least one vertex $u\in C' \setminus \mathcal{N}(C')$ must survive after deleting the vertices in a solution. This vertex must be part of a clique of size $c$. Therefore, the clique $C'$ must appear unchanged in the resulting graph. Hence, we must delete all the neighbours of $C'$. \qed \\

\noindent Let us denote by $\mathcal{L}^{\prime}_{iso,c}$ the set of isolated cliques of size $c$ in $\mathcal{L}^{\prime}$ obtained after applying Reduction Rule \ref{Red8}. 
Similarly,  denote by $\mathcal{L}^{\prime}_{>c}$ the set of cliques of size more than $c$ in $\mathcal{L}^{\prime}$.
Note that, we have  $\mathcal{L}^{\prime} = \mathcal{L}^{\prime}_{iso,c} \cup \mathcal{L}^{\prime}_{>c}$.

\begin{observation}
    Due to application of Reduction Rule \ref{Red2}, we get $|\mathcal{L}^{\prime}_{iso,c}|\leq k+1$. 
\end{observation}

\begin{evd}\label{Red10}
    If $|\mathcal{L}^{\prime}_{>c}|>k$, then conclude that we are dealing with a 
    no-instance.
\end{evd}

\noindent If $|\mathcal{L}^{\prime}_{>c}|>k$,  it implies we need to delete
more than $k$ vertices to ensure that each clique in $\mathcal{L}^{\prime}_{>c}$
becomes a clique of size $c$ in the resulting graph. However, since we are constrained to finding a
solution of size at most $k$, achieving this is not feasible. 
 Therefore, $|\mathcal{L}^{\prime}_{>c}|\leq k$.

\begin{observation}\label{obs3}
    After applying Reduction Rule \ref{Red7}, \ref{Red8} and \ref{Red10}, the 
    size of $\mathcal{L}^{\prime}$ is bounded by $2k+1$.
\end{observation}
We know that $\mathcal{L}^{\prime} = \mathcal{L}^{\prime}_{iso,c} \cup \mathcal{L}^{\prime}_{>c}$. Therefore,  
$|\mathcal{L}^{\prime}|\leq |\mathcal{L}^{\prime}_{iso,c}| + |\mathcal{L}^{\prime}_{>c}| \leq (k+1)+k \leq 2k+1$.
\vspace{3mm}

\noindent Clearly, due to Observation \ref{obs3}, we have $\bigcup\limits_{C \in \mathcal{L}^{\prime}} |C\setminus \mathcal{N}(C)| \leq \mathcal{O}(k^{3})$. 
This completes the proof of Theorem \ref{UCVD}.

\begin{theorem}\label{FPTevd}
     The {\sc UCVD} problem parameterized by solution size $k$ can be solved in $2^{k}\cdot n^{\mathcal{O}(1)}$ time.
\end{theorem}
\proof We utilize an algorithm from \cite{TsurD} to find a cluster vertex deletion set $X$ of size
at most $k$ in  $\mathcal{O}^{*}(1.811^{k})$ time. If this algorithm determines that 
no such set exists, conclude that we are dealing with a no-instance. 
Otherwise, this algorithm returns a cluster vertex deletion set $X$ of size at most $k$.  To decide whether $G$ contains a uniform cluster vertex deletion set $S$ of size 
at most $k$, we proceed as follows: 
\begin{itemize}
    \item We guess the intersection of $S$ with $X$, that is, 
we guess the set $X_{in}=X\cap S$, delete $X_{in}$ from $G$ and reduce
parameter $k$ by $|X_{in}|$.
\item For each guess of $X_{in}$, we set 
$X_{out}= X\setminus X_{in}$ and solve {\sc Disjoint UCVD} on the instance 
$(G-X_{in}, X_{out}, k-|X_{in}|)$.
\end{itemize}
For each guess $X_{in}$, we try to find a  uniform 
cluster vertex deletion set $S'$ in $G-X_{in}$ of size at most 
$k-|X_{in}|$ that is disjoint from $X_{out}$. This problem is called
{\sc Disjoint UCVD}.
If for some guess $X_{in}$, we find a uniform 
cluster vertex deletion set $S'$ in $G-X_{in}$ of size at most 
$k-|X_{in}|$ that is disjoint from $X_{out}$, then we output $S=X_{in} \cup S'$.
Otherwise, we conclude that the given instance of the {\sc Disjoint UCVD} problem is
a no-instance.  The number of all guesses is bounded by $2^k$.
Therefore, to obtain an FPT 
algorithm for {\sc UCVD}, it is sufficient to solve {\sc Disjoint UCVD} in polynomial time. \\

\noindent \textbf{An algorithm for {\sc Disjoint UCVD:}}
Let $(G-X_{in}, X_{out},k)$ be an instance of {\sc Disjoint UCVD}, and let $H=G-X$,
where $X=X_{in}\cup X_{out}$. We denote $G'=G-X_{in}$.

\begin{disevd}\label{xout}
    If $G[X_{out}]$ is not a disjoint union of cliques  or $G-X$ is not a disjoint union of cliques, then return that $(G', X_{out},k)$ is a no-instance.
\end{disevd}

\begin{disevd}\label{p3}
   If there is a vertex $v$ in $G-X$ which has neighbours in two distinct cliques in $G[X_{out}]$, then delete $v$ from $G$ and decrement the parameter $k$ by 1. The new instance is $(G'-\{v\},X_{out},k-1)$.
\end{disevd}

\noindent The previous reduction rule is correct because we will get an induced $P_3$ where the only vertex that we are allowed to delete is $v$. Therefore, we must add $v$ to the solution. \\

\noindent So from now onwards, we assume that $G[X_{out}]$ is  a disjoint union of cliques  and $H=G-X$ is  a disjoint union of cliques.  We also guess the exact size of the solution as $k'$. 
The possible guesses for $k'$ range from $1$ to $k$. This will  be helpful in the final part of the algorithm.
We will solve this problem by guessing the size $c$ of cliques in the resulting graph.
At the end, we will compare the solution sizes for each $c$ and return the minimum one.
The possible guesses for $c$ ranges from $1$ to $n$. 

\begin{disevd}\label{csizeclique}
    If there is a clique of size exactly $c$ in $G[X_{out}]$ then delete $N(C)$ and add all the vertices of the set $N(C)$ to the solution. The new instance is $(G'-N[C],X_{out},k-|N(C)|)$.
\end{disevd}


\begin{observation}\label{largesizeclique}
    If there is a clique $C$ of size larger than $c$ in $G[X_{out}]$, then we can discard such a guess for $c$.
\end{observation}

\begin{observation}\label{lessthanc}
    Every clique in $G[X_{out}]$ of size less than $c$ must utilize vertices from exactly one clique in $G-X$ to form a clique of size $c$.  
\end{observation}

\noindent Due to Observation \ref{lessthanc}, for every clique $C$ of size less than $c$ in $G[X_{out}]$, there is a unique clique in $G-X$ that helps to form a clique of size $c$ containing vertices from $C$. 

\par We now construct a bipartite graph with bipartition $(A,B)$ in the following way. 
For a given guess $(k',c)$, observe that the number of equal-sized cliques in 
the resulting graph is $p=\frac{|V(G')|-k'}{c}$. We need a total of $p$ vertices 
in $A$. Suppose the number of cliques in $G[X_{out}]$ is $p_1$. We add a vertex $a\in A$ corresponding  to 
each clique in $G[X_{out}]$. If $p_1<p$, 
we add $p-p_1$
dummy  vertices in $A$ to ensure the number of vertices in $A$ equals $p$. 
Similarly, we add a vertex   $b\in B$ corresponding  to each clique in $G-X$.
Due to Reduction Rule Disjoint-UCVD \ref{csizeclique} and Observation \ref{largesizeclique},
it follows that every clique in $G[X_{out}]$ has size less than $c$.
We add an edge between $a\in A$ and $b\in B$ if the clique corresponding to $a$ 
can be transformed into a clique of size $c$ using some vertices from the clique  corresponding to $b$. 
Additionally, some cliques in $G-X$ can be transformed into cliques of size $c$ by deleting 
the necessary number of vertices from the respective cliques. 
We make $b\in B$ adjacent to all dummy vertices in $A$  if the clique 
corresponding to $b$ can be transformed into a clique of size $c$ by deleting
the necessary number of vertices. 

\begin{lemma}
    There exists an $A$-saturated matching if and only if there is a set of $p$ disjoint cliques, each of size exactly $c$, obtained by deleting exactly $k'$ vertices from $G-X$, such that the graph induced by these cliques forms a uniform cluster graph.
\end{lemma}

\proof In the forward direction, suppose there exists an $A$-saturated matching. We can identify $p$ cliques in the resulting graph as follows:

\noindent - If $(a,b) \in M$ and $a$ is not a dummy vertex, the clique corresponding to $a \in A$ will be transformed into a clique of size $c$ using some vertices from the clique corresponding to $b \in B$.
- If $(a,b) \in M$ and $a$ is a dummy vertex, the clique corresponding to $b \in B$ will be adjusted to a clique of size $c$ by deleting the necessary vertices from that clique.

Thus, we can identify the exact cliques in the final graph. Since $p \cdot c = |V(G')| - k'$, we know exactly which $k'$ vertices need to be deleted to form these cliques of size $c$.

In the backward direction, assume there is a set of $p$ disjoint cliques, each of size exactly $c$, obtained by deleting exactly $k'$ vertices from $G-X$, and that the induced subgraph on these cliques forms a uniform cluster graph. By our construction, we can obtain an edge in the bipartite graph $(A,B)$ corresponding to each clique, forming a matching. Since there are $p$ cliques, this matching will be $A$-saturated. \qed

\noindent We now try to find an $A$-saturated matching in this bipartite  graph. This can be done in polynomial time.
Due to the previous claim, if an $A$-saturated matching does not exist, we discard such a guess $(k',c)$. 
Otherwise, we find an $A$-saturated matching $M$ of size $p$ and return a yes instance. This finishes the proof of Theorem \ref{FPTevd}. 

\section{Kernelization algorithm for {\sc UCVS} parameterized by solution size}\label{Section UCVS}

Suppose $G$ is a graph with an isolated vertex. If $(G,k)$ is a yes-instance then the size of equal-sized cliques in the resulting graph must be 1. Therefore if $(G,k)$ is yes-instance then $G$ itself is a disjoint union of isolated vertices. We can solve $(G,k)$ in polynomial time if $G$ has an isolated vertex.
So in this section, all the graphs considered are without any isolated vertex.

In this section, we prove that {\sc UCIVS} and {\sc UCEVS} are NP-hard. We provide a reduction from a problem called {\sc $K_d$-Edge Partition}.

\begin{definition}
    A graph \(G = (V, E)\) is said to have a \(K_{d}\)-edge partition if there exists a partition of the edge set \(E\) into subsets \(\mathcal{P} = \{P_1, P_2, \dots, P_l\}\), such that the subgraph \(G[P_i]\) induced by the edges in each \(P_i\) is isomorphic to a clique of size \(d\).
\end{definition}

\noindent The following problem was shown to be NP-complete by Ian Holyer\cite{doi:10.1137/0210054} for each $d\geq 3$.

\vspace{3mm}

\fbox
    {\begin{minipage}{33.7em}\label{USCC}
       {\sc $K_d$-Edge Partition}\\
        \noindent{\bf Input:} An undirected graph $G$.\\
    \noindent{\bf Question:} Is there a $K_d$-Edge Partition $\mathcal{P}$ of $G$?
    \end{minipage} }\\
\vspace{3mm}

\noindent To make the explanation easier, we further define cost of a partition.

\begin{definition}
Let \(\mathcal{P}\) be a \(K_d\)-Edge Partition of \(G\). The cost of \(\mathcal{P}\), denoted by \(\text{cost}(\mathcal{P})\), is defined as:
\[
\text{cost}(\mathcal{P}) = \sum_{v \in V} \max(0,\mathrm{freq}_{\mathcal{P}}(v)-1),
\]
where $\mathrm{freq}_{\mathcal{P}}(v) := |\{P \in \mathcal{P} \mid v \in V(G[P])\}|$.
\end{definition}

\vspace{3mm}
\fbox
    {\begin{minipage}{33.7em}\label{}
       {\sc Weighted $K_d$-Edge Partition}\\
        \noindent{\bf Input:} An undirected graph $G=(V,E)$, and a positive integer $k$.\\
    \noindent{\bf Question:} Is there a $K_d$-Edge Partition $\mathcal{P}$ of $G$ with $\texttt{cost}(\mathcal{P}) \leq k$?
    \end{minipage} }\\
\vspace{3mm}

\noindent We include a few lemmas to simplify and clarify the explanation of the NP-hardness proof.

\begin{lemma}\label{esplitting}
    
Let $G=(V, E)$ be a graph without an isolated vertex and let $\mathcal{P}$ be a $K_{d}$-Edge Partition of $G$ with $\text{cost}_G(\mathcal{P}) = k$. Then there is a vertex $u \in V$ such that by doing an exclusive vertex splitting of $u$ in $G$, we obtain $G_{1}=\left( V_{1}, E_{1} \right)$ satisfying 
\begin{enumerate}
    \item  $G_{1}$ has a $K_{d}$-Edge Partition, say $\mathcal{P}^\prime$,
    \item  $\text{cost}_{G_{1}}(\mathcal{P}^\prime)=k-1$.
\end{enumerate}

\end{lemma} 

\proof If $G$ is not already a disjoint union of $d$-sized cliques, without loss of generality there must exist $P_1$ and $P_2$ such that $V(G[P_{1}]) \cap V(G[P_{2}])  \neq \emptyset$. In this case, let $u \in V(G[P_{1}]) \cap V(G[P_{2}])$.
We define $G'$ as the graph that is obtained when $u$ is split into the two vertices $u_{1}$ and $u_{2}$ such that 
$$
\begin{gathered}
N_{G_{1}}(u_{1}):=N_G(u) \cap V(G[P_{1}]), \\
N_{G_{1}}(u_{2}):=N_G(u) \setminus V(G[P_{1}]).
\end{gathered}
$$

\noindent Furthermore, we obtain a $K_{d}$-edge partition of $G_{1}$ as \(\mathcal{P} = \{P'_1, P'_2, \dots, P'_l\}\) such that 

$$
P'_i:= \begin{cases} \{\{x, y\} \in P_1 \mid u \notin \{x, y\}\} \cup \{\{u_1, v\} \mid \{u, v\} \in P_1\} & \text { if } i=1
\\  \{\{x, y\} \in P_i \mid u \notin \{x, y\}\} \cup \{\{u_2, v\} \mid \{u, v\} \in P_i\} & \text { if } 1 < i \leq l\end{cases}
$$  

\noindent Note that $\mathrm{freq}_{\mathcal{P}}(u)=\mathrm{freq}_{\mathcal{P'}}(u_{1})+\mathrm{freq}_{\mathcal{P'}}(u_{2})$. In particular, we have $\mathrm{freq}_{\mathcal{P'}}(u_{1})=1$ and $\mathrm{freq}_{\mathcal{P'}}(u_{2})=\mathrm{freq}_{\mathcal{P'}}(u)-1$. Therefore, we get $\text{cost}_{G_{1}}(\mathcal{P}^\prime)=k-1$.  \qed

\begin{lemma}\label{equi ucevs and partition}
    $(G,k)$ is a yes-instance of {\sc UCEVS} if and only if $(G,k)$ is a yes-instance of {\sc Weighted $K_{d}$-edge partition} for some $d$.
\end{lemma}
\proof Let $I=(G,k)$ be a yes instance, that is, there exists a sequence of at most $k$ vertex splittings in $G$ such that the resulting graph $G'$ is a cluster graph containing equal sized cliques of size $d$.
Since we are doing exclusive vertex splitting, for every edge $u'v'\in E(G')$, there exists an unique edge $uv$ in $E(G)$ where either $u'=u$ or $u'$ is a descendant of $u$ and similarly $v'=v$ or $v'$ is a descendant of $v$.
So, for every $K_{d}$-sized cliques in $G'$ there exists a unique $K_{d}$-sized clique in $G$. 
So we get a $K_{d}$-Edge partition of $G$. 
Since we are doing at most $k$ number of vertex splitting, the number of vertices in $G'$ is at most $|V(G)|+k$. 
Hence $(G,k)$ is a yes-instance of {\sc Weighted $K_{d}$-Edge Partition}.

For the reverse direction, let $\mathcal{P}$ be a $K_{d}$-Edge Partition of $G$ with $\texttt{cost}(\mathcal{P}) \leq k$.
By iteratively applying Lemma \ref{esplitting} $k$ times, there exists a sequence of exclusive vertex splittings of length at most $k$ such that we obtain a disjoint union of $d$-sized cliques.
\qed

\begin{theorem}\label{NP-hard UCIVS}
        {\sc Uniform Cluster Inclusive Vertex Splitting} is NP-complete.
\end{theorem}
\proof It is clear that the problem is in NP. 
In order to show that the problem is NP-hard, we provide a reduction from {\sc $K_{3}$-edge partition}. 
The problem asks to determine whether the given graph
can be edge-partitioned into subgraphs isomorphic to the complete graph $K_{3}$.
Ian Holyer proved that this problem is NP-hard \cite{doi:10.1137/0210054}.
Given an instance $I=G(V,E)$ of {\sc $K_{3}$-edge partition}, we construct an instance $I'=(G',k')$ of {\sc Uniform Cluster Inclusive Vertex Splitting}.
To construct the graph $G'$, we start with a graph $G$ and add $m-n+1$ disjoint triangles to it where $|V(G)|=n$ and $|E(G)|=m$.
We set $k'=m-n$. 
Next, we prove that the instances $I$ and $I'$ are equivalent.

Let $I$ be a yes instance, that is, there exists a partition, say $\mathcal{P}$ of edges of $G$ such that the graph induced by edges in each part is isomorphic to $K_{3}$ (a triangle). 
We denote such a partition as
$\mathcal{P} = \{ P_1, P_2, \dots, P_{\frac{m}{3}} \mid G[P_i] \cong K_3, \, \forall i \in \{1, 2, \dots, {\frac{m}{3}}\} \}$.
If such a partition exists then each vertex in $G$ must have an even degree.
For each vertex $v\in V(G)$, $v$ is part of $l$ number of elements of $\mathcal{P}$ if and only if $d(v)=2l$. Therefore $\texttt{cost}(\mathcal{P})= \sum_{v \in V(G)} \bigg( \frac{d(v)}{2} - 1 \bigg)= m - n$. Now we apply iteratively Lemma \ref{esplitting} by taking $K_{d}=K_3$ until we get a disjoint union of triangles. Note that the total number of vertex splitting required is equal to 
$ \sum_{v \in V(G)} \bigg( \frac{d(v)}{2} - 1 \bigg)= m - n$.

Let $I'$ be a yes instance, that is, there exists a sequence of at most $m-n$ vertex splittings in $G'$ such that the resulting graph $G''$ is uniform cluster graph. 
We first argue that $G''$ is a disjoint union of cliques of size 3.
Note that this must be true as $G'$ contains $m-n+1$ many connected components which are triangles and since only $m-n$ vertex splittings are allowed, $G''$ contains at least one disjoint triangle. 
As $G''$ is uniform cluster graph, it must be a disjoint union of cliques of size 3.
We can assume that no vertices in the set $V(G')\setminus V(G)$ are split.

\begin{claim}
     For every edge $u''v''\in E(G'')$, there exists a unique edge $uv$ in $E(G')$ where either $u''=u$ or $u''$ is obtained from splitting of a vertex $u$ and similarly $v''=v$ or $v''$ is obtained from splitting of a vertex $v$.
\end{claim}
\proof Note that after $m-n$ vertex splittings, the graph $G''$ can contain at most $n+3(m-n+1)+(m-n)=m+3(m-n+1)$ many vertices. Since $G''$ is disjoint union of cliques of size 3, $E(G'') = \frac{3|V(G'')|}{3}  = V(G'') \leq m + 3(m-n+1) = E(G')$. Since $G''$ is obtained by vertex splitting operations on $G'$, we must have $E(G'') \geq E(G')$. This implies that $E(G'')=E(G')$. 
Therefore, for every edge $u''v''\in E(G'')$, there exists an unique edge $uv$ in $E(G')$. \claimqed
\vspace{3mm}
\noindent The above claim shows that, for every  triangle in $G''$ there exists a unique corresponding triangle in $G'$. 
As $|E(G'')|=|E(G')|$, this gives us a partition of edges of $G'$ where each part induces a $K_{3}$. Given a partition of $G'$, one can easily get a partition of edges of $G$ as well. This completes the proof of \ref{NP-hard UCIVS}. \qed 
\vspace{3mm}

\noindent Note that the above proof also works for exclusive vertex splitting as well as one can see that in the backward part of the proof, all the vertex splitting done are exclusive vertex splittings.

\begin{corollary}\label{NP-hard UCEVS}
        {\sc Uniform Cluster Exclusive Vertex Splitting} is NP-complete.
\end{corollary}

\subsection{A Kernelization Algorithm for {\sc UCEVS}}

Now, we will present a linear vertex kernel for the {\sc UCEVS}.

\begin{theorem}\label{UCEVSkernel}
    {\sc Uniform Cluster-Exclusive Vertex Splitting} parameterized by solution size admits a kernel with at most $4k$ vertices.
\end{theorem}

\proof Let $G_0,G_1,G_2,...G_l$ be a splitting sequence with size $l$ for $G$. Let $S = \{v_1,v_2,\ldots,v_l\}$ be the collection of vertices that were split to get $G_l$ from $G_0=G$. We observe that if there is indeed a split sequence of size at most $k$, then the vertices in  $V\setminus S$ have equal degrees in both $G$ and $G_l$. 
In other words, for every $x \in V\setminus S$, we have $d_{G}(x)=d_{G_l}(x)=d$ for some $d$. We assume that $G$ has at least $2k+1$ vertices; otherwise, we have a kernel of size $2k$.
We know that if the input is a yes-instance, then all but $k$ vertices have the same degree $d$. Therefore, we first check if the input instance $(G,k)$ satisfies this condition. If the input instance does not satisfy this condition, then conclude that we are dealing with a no-instance. If the input instance satisfies this condition, we can calculate the exact value of $d$ in linear time. Assuming that the input is a yes-instance, let $c$ be the size of equal-sized cliques in $G_l$.
One can observe that $c$ must be equal to $d+1$ in $G_l$.

Since the graph does not have isolated vertices, every vertex must be part of a clique in any $K_{d+1}$-Edge Partition.

\begin{ucevs}\label{UCEVS_combined}
    If any of the following conditions hold, conclude that we are dealing with a no-instance:
    \begin{enumerate}
        \item \( |\{v \in V(G) \mid d_G(v) > d\}| > k \),
        \item There exists a vertex \( v \in V(G) \) such that \( d_G(v) < d \),
        \item There exists a vertex \( v \in V(G) \) such that \( d_G(v) = d \) and \( G[N[v]] \) is not a clique.
    \end{enumerate}
\end{ucevs}

\begin{lemma}\label{RR1 evs combined correct}
    Reduction Rule \ref{UCEVS_combined} is correct.
\end{lemma}
\begin{proof}
We justify the correctness of each condition in Reduction Rule \ref{UCEVS_combined} as follows:

\begin{enumerate}
    \item \textbf{Condition 1}: If \( |\{v \in V(G) \mid d_G(v) > d\}| > k \), then there are at least \( k+1 \) vertices whose frequency must be at least two in any $K_{d+1}$-Edge Partition. This means for any $K_{d+1}$-Edge Partition of $G$ cost must be greater than $k$. Therefore, the instance must be a no-instance.


    \item \textbf{Condition 2}: If there exists a vertex \( v \in V(G) \) such that \( d_G(v) < d \), then clearly $G[N[v]]$ does not contain a clique of size $d+1$. It means $v$ can not be part of a clique in a $K_{d+1}$-Edge Partition. Thus, the instance is a no-instance. 
    

    \item \textbf{Condition 3}: If there exists a vertex \( v \in V(G) \) such that \( d_G(v)=d \) and $G[N[v]]$ is not a clique. It means $G[N[v]]$ does not contain a clique of size $d+1$ and $v$ can not be part of a clique in a $K_{d+1}$-Edge Partition. Thus, the instance is a no-instance.
    
    
\end{enumerate}

Since all three conditions correctly identify invalid instances, the reduction rule is sound.  
\end{proof}

\noindent We make two cases based on the value of $d$.

\vspace{3mm}

\noindent {\bf Case 1.} Let $d>2k$.

\noindent In this case, we provide a polynomial time algorithm to solve the problem, that is, due to Lemma \ref{equi ucevs and partition}, we determine whether there exists a weighted $K_{d+1}$-edge partition of $G$ with cost at most $k$.

\begin{tcolorbox}[colframe=black, colback=white, boxrule=0.5mm, title={Algorithm: Weighted $K_{d+1}$-Edge Partition}]
\begin{algorithmic}[1]\label{Edge Partition}
\Require A graph \( G = (V, E) \) and an integer \( k \).
\Ensure A weighted \( K_{d+1} \)-edge partition \(\mathcal{P}\) of cost at most \( k \), or conclude that no such partition exists.
\State Initialize an empty partition: \(\mathcal{P} \gets \emptyset\)
\While{\( G \neq \emptyset \)}
    \State Find a vertex \( v \in V(G) \) such that \( d_G(v) = d \)
    \If{no such vertex exists}
        \State \Return "No instance."
    \EndIf
    \State Let \( C = G[N[v]] \), the subgraph induced by the closed neighborhood of \( v \)
    \If{\( C \) is not a clique}
        \State \Return "No instance."
    \EndIf
    \State Add the edges of \( C \) to \(\mathcal{P}\):
    \[
    \mathcal{P} \gets \mathcal{P} \cup \{E(C)\}
    \]

    \State \textbf{Check Consistency of \(\mathcal{P}\):}
    \begin{itemize}
        \item If any two elements \( P_i, P_j \in \mathcal{P} \) share an edge (i.e., \( P_i \cap P_j \neq \emptyset \)), return "No instance."
        \item Compute \(\text{cost}(\mathcal{P}) =  \max(0,\mathrm{freq}_{\mathcal{P}}(v)-1)\). If \(\text{cost}(\mathcal{P}) > k\), return "No instance."
    \end{itemize}

    \State \textbf{Update \( G \) and \( k \):}
    \begin{align*}
    V(G) &\gets V(G) \setminus \{u \in N[v] \mid d_G(u) = d\}, \\
    k &\gets k - \mathrm{cost(\mathcal{P})}
    \end{align*}
\EndWhile
\State \Return \(\mathcal{P}\)
\end{algorithmic}
\end{tcolorbox}

\begin{lemma}\label{correctness of partition algorithm1}
    The Weighted \( K_{d+1} \)-Edge Partition Algorithm (Algorithm~\ref{Edge Partition}) correctly outputs a weighted $K_{d+1}$-edge partition of cost at most $k$, or conclude that no such partition exists.
\end{lemma}
\begin{proof}
The correctness of the algorithm is established as follows:

1. \textbf{Initial If Conditions}:  
   The first two \texttt{if} conditions (lines 4 and 7 of the algorithm) directly follow from \textbf{Reduction Rule \ref{UCEVS_combined}}:
   \begin{itemize}
       \item If no vertex \( v \) exists with \( d_G(v) = d \), then it is impossible to construct a valid \( K_{d+1} \)-edge partition, as per \textbf{Condition 1} of the reduction rule.
       \item If \( G[N[v]] \) is not a clique for a vertex \( v \) with \( d_G(v) = d \), then \( v \) cannot belong to any \( K_{d+1} \)-clique, as per \textbf{Condition 3} of the reduction rule.
   \end{itemize}

2. \textbf{Partition Update}:  
   When a vertex \( v \) with \( d_G(v) = d \) is found, all edges adjacent to \( v \) must belong to the same part of the partition \(\mathcal{P}\). This is guaranteed by the definition of a \( K_{d+1} \)-edge partition, which requires that \( G[N[v]] \) form a clique.  
   Therefore, adding \( E(G[N[v]]) \) to \(\mathcal{P}\) is a valid and safe update.

3. \textbf{Consistency Checks}:  
   \begin{itemize}
       \item The \textbf{first consistency condition} holds because \(\mathcal{P}\) is explicitly defined as a partition of the edge set, ensuring that no two parts overlap.
       \item The \textbf{second consistency condition} holds due to \textbf{Lemma \ref{equi ucevs and partition}}, which guarantees that cost of the partition \(\mathrm{cost}(\mathcal{P})\) does not exceed the allowed budget \( k \).
   \end{itemize}

4. \textbf{Graph Updates}:  
   Since all edges adjacent to the vertices in the set \(\{u \in N[v] \mid d_G(u) = d\}\) have been included in \(\mathcal{P}\), it is safe to delete these vertices and their incident edges from \( G \).  
   Additionally, the budget \( k \) is correctly reduced by the amount \(\mathrm{cost}(\mathcal{P})\), reflecting the edges already accounted for in the partition.

5. \textbf{Termination}:  
   The algorithm terminates when \( G \) becomes empty, at which point all edges have been successfully partitioned into \( K_{d+1} \)-cliques. If any condition fails during execution, the algorithm correctly concludes a no-instance.
Since $d\geq 2k$, every vertex of degree greater than $d$ is adjacent to a vertex of degree $d$ otherwise Reduction Rule UCIVS \ref{UCEVS_combined} could have been applied. Note that after each iteration, if $G$ is not empty, we will either have a vertex of degree $d$ or all vertices have a degree not equal to $d$. In the latter case, we return no-instance by Reduction Rule \ref{UCEVS_combined} and in the former case, since we have a vertex with degree $d$, we execute the line 3 of the algorithm.  This means that either $G$ becomes empty or any condition fails during execution. 

Thus, the algorithm is correct.
\end{proof}

\noindent \textbf{Case 2}. Let us assume $d \leq 2k$.

There is a subtle difference between the two cases. In the case 1, after each iteration of updating \( G \), one of two outcomes occurs. Either the graph \( G \) becomes empty, or \( |V(G)| \geq 2k+1 \). In the former case, we can output the partition directly. In the latter case, Reduction Rule \ref{UCEVS_combined} guarantees that at least \( k+1 \) vertices (and all but \( k \)) have degree \( d \). This ensures that if the reduced input is a yes instance, the equal-sized cliques in the final graph must be of size \( d+1 \), which is consistent with the original instance.

On the other hand, in case 2, we must ensure that after each update of \( G \), either the graph becomes empty or it contains at least \( 2k+1 \) vertices. In this case, Reduction Rule \ref{UCEVS_combined} again ensures that at least \( k+1 \) vertices (and all but \( k \)) have degree \( d \). Since \( d \leq 2k \), the graph must have at least \( 4k+1 \) vertices to satisfy these conditions. Therefore, we apply the Weighted \( K_{d+1} \)-Edge Partition Algorithm (Algorithm~\ref{Edge Partition}) only if \( |V(G)| > 4k \).

\begin{ucevs}\label{UCEVS_d_leq_2k}
    If \( |V(G)| > 4k \), then run the Weighted \( K_{d+1} \)-Edge Partition Algorithm (Algorithm~\ref{Edge Partition}) on the graph \( G \), but terminate the algorithm if the size of the vertex set becomes \( |V(G)| \leq 4k \).
\end{ucevs}

\noindent The correctness of the Reduction Rule \ref{UCEVS_d_leq_2k} follows from correctness of Weighted \( K_{d+1} \)-Edge Partition Algorithm (Algorithm~\ref{Edge Partition}) due to Lemma \ref{correctness of partition algorithm1}.
This finishes the proof of Theorem \ref{UCEVSkernel}. \qed

\subsection{A Kernelization Algorithm for {\sc UCIVS}}

We will present a linear kernel for the UCIVS.

\begin{theorem}\label{UCIVSkernel}
    {\sc Uniform Cluster-Exclusive Vertex Splitting} parameterized by solution size admits a kernel with at most $4k$ vertices.
\end{theorem}

\noindent We first provide some lemmas given by Firbas et.al \cite{10.1007/978-3-031-52113-3_16}.

\begin{lemma}\label{noincident}\cite{10.1007/978-3-031-52113-3_16}
    Let $G$ be a non-empty graph and $G'$ be obtained from $G$ by splitting some vertex $w \in V (G)$  into $w_1, w_2 \in V (G')$; furthermore let $v_1, v_2 \in V (G)$. If $v_1v_2 \notin E(G)$,
then for all descendants $v'_ 1, v'_ 2 \in V (G')$ of $v_1$ and $v_2$ respectively, it holds that $v'_1 v'_2 \notin E(G')$.
\end{lemma}

\begin{lemma}\label{incident}\cite{10.1007/978-3-031-52113-3_16}
    Let $G$ be a non-empty graph and $G'$ be obtained from $G$ by splitting some vertex $w \in V (G)$  into $w_1, w_2 \in V (G')$; furthermore let $v_1, v_2 \in V (G)$. If $v_1v_2 \in E(G)$,
then there are two descendants $v'_ 1, v'_ 2 \in V (G')$ of $v_1$ and $v_2$ respectively, such that $v'_1 v'_2 \in E(G')$.
\end{lemma}

\noindent Note that in the previous subsection, we saw correlation between the problems {\sc UCEVS} and {\sc $K_{d+1}$-Edge Partition}. Here, we see a similar trend. We correlate the {\sc UCIVS} and a problem called {\sc Sigma uniform clique cover  (SUCC)}.

\begin{definition}\label{sclique}\rm 
    A family $\mathcal{C}$ of cliques in $G$ is called an \emph{Sigma uniform clique cover } if
    \begin{enumerate}
        \item for each $e \in E$, there exists $C \in \mathcal{C}$ such that $e \in E(G[C])$ and
        \item all the cliques in the family  $\mathcal{C}$ have the same size. 
    \end{enumerate}
\end{definition}

\begin{definition}
    A Sigma uniform clique cover where cliques are of size $s$ is called a Sigma $s$-clique cover.
\end{definition}

\begin{definition}
Let \(\mathcal{C}\) be a Sigma uniform clique cover of \(G\). The weight of \(\mathcal{C}\), denoted by \(\texttt{wgt}(\mathcal{C})\), is defined as:
\[
\texttt{wgt}(\mathcal{C}) = \sum_{v \in V} \mathrm{freq}_{\mathcal{C}}(v),
\]
where $\mathrm{freq}_{\mathcal{C}}(v) := |\{C \in \mathcal{C} \mid v \in V(G[C])\}|$.
\end{definition}

\noindent A similar definition of weight of $\mathcal{C}$ is given in \cite{10.1007/978-3-031-52113-3_16}. We also define a cost of $\mathcal{C}$. The definition of cost of $\mathcal{C}$ will be useful in the explanation of kernelization result.

\begin{definition}
Let \(\mathcal{C}\) be a Sigma uniform clique cover of \(G\). The cost of \(\mathcal{C}\), denoted by \(\text{cost}(\mathcal{C})\), is defined as:
\[
\text{cost}(\mathcal{C}) = \sum_{v \in V}  \max(0,\mathrm{freq}_{\mathcal{C}}(v)-1),
\]
where $\mathrm{freq}_{\mathcal{C}}(v) := |\{C \in \mathcal{C} \mid v \in V(G[C])\}|$.
\end{definition}

The following two lemmas are closely related to lemmas proved in  \cite{10.1007/978-3-031-52113-3_16}. Lemmas given in \cite{10.1007/978-3-031-52113-3_16} dealt with the problem {\sc Cluster Inclusive Vertex Splitting} relating it to {\sc sigma clique cover}. Here, we see that the same idea works while working with the {\sc UCIVS} as well by taking {\sc sigma uniform clique cover} instead of {\sc sigma clique cover}.

\begin{lemma}\label{merging}
    Let $G=(V, E)$ be a graph and let $G'=(V', E')$ be obtained from $G$ by splitting $u \in V$ into $v, w \in V'$. If $\mathcal{C}'$ is a Sigma uniform clique cover  of $G'$, then there exists a Sigma uniform clique cover  $\mathcal{C}$ of $G$ with \texttt{wgt}$(\mathcal{C}) = $\texttt{wgt}$(\mathcal{C}')$.
\end{lemma} 

\proof Using
$$
f(C'):= \begin{cases}(C' \setminus \{v, w\}) \cup\{u\} & \text { if } C' \cap\{v, w\} \neq \emptyset \\ C' & \text { otherwise }\end{cases}
$$
we define
$$
\mathcal{C}:=\left\{f\left(C'\right) \mid C' \in \mathcal{C}'\right\} .
$$

Note that $f$ gives a bijection over the domain $\mathcal{C}'$. We claim that $\mathcal{C}$ satisfies the conditions of this lemma. First, we establish that $\mathcal{C}$ is a Sigma uniform clique cover  of $G$. We note that the map $f$ is the same as the map given in the proof of lemma $4.1$ of \cite{10.1007/978-3-031-52113-3_16}. 
So, we can conclude that $C \in \mathcal{C}$ induce cliques in $G$  and all edges of $G$ are covered by $\mathcal{C}$. If $C' \cap \{v,w\} \neq \emptyset$, then since $C'$ is a clique, it can not have both $v$ and $w$, it means $C'$ contains exactly one vertex from $v$ and $w$. Therefore $|f(C')|=|C'|$ when $C' \cap \{v,w\} \neq \emptyset$. It is clear that $|f(C')|=|C'|$ when $C' \cap \{v,w\} = \emptyset$. Hence $f$ ranging over $\mathcal{C}'$ does not change the cardinality of any image it maps, this observation says that all cliques in $\mathcal{C}$ are of the same size. It is also clear that $\texttt{wgt}(\mathcal{C})=\texttt{wgt}\left(\mathcal{C}'\right)$.
\qed

\begin{lemma}\label{splitting}
    
Let $G=(V, E)$ be a graph with no isolated vertices and let $\mathcal{C}$ be a Sigma uniform clique cover  of $G$ with $\operatorname{wgt}(\mathcal{C}) \leq|V|+\alpha \in \mathbb{N}$. Then, either $G$ is already a uniform cluster graph or there is $u \in V$ such that $u$ can be split in $G$ to obtain $G'=\left(V', E'\right)$ satisfying 
\begin{enumerate}
    \item  $G'$ has a Sigma uniform clique cover  $\mathcal{C}'$,
    \item  $\operatorname{wgt}\left(\mathcal{C}'\right) \leq\left|V'\right|+\alpha-1$,
    \item  $G'$ does not contain isolated vertices.
\end{enumerate}

\end{lemma} 

Proof. If $G$ is not already a cluster graph, there must exist $C_1 \neq C_2 \in \mathcal{C}$ such that $C_1 \cap C_2 \neq \emptyset$. In this case, let $u \in C_1 \cap C_2$.

We define $G'=(V', E')$ as the graph that is obtained when $u$ is split into the two vertices $u_{\text {in}}$ and $u_{\text {out}}$ obeying:
$$
\begin{gathered}
N_{G'}(u_{\text {in }}):=N_G(u) \cap C_1, \\
N_{G'}(u_{\text {out }}):=(N_G(u) \setminus C_1) \cup \{v \in N_G(u) \cap C_1 \mid \exists C \in \mathcal{C} \setminus\left\{C_1\right\}: u, v \in C\} .
\end{gathered}
$$

Furthermore, using the map
$$
f(C):= \begin{cases}(C \setminus\{u\}) \cup\{u_{\text {in}}\} & \text { if } C=C_1 \\ (C \setminus\{u\}) \cup\{u_{\text {out}}\} & \text { if } u \in C \wedge C \neq C_1 \\ C & \text { otherwise }\end{cases}
$$
we can define
$$
\mathcal{C}':=\{f(C) \mid C \in \mathcal{C}\} .
$$

Note that $f$ gives a bijection between $\mathcal{C}$ and $\mathcal{C}'$. 
Thus, $\mathcal{C}'$ is a sigma clique cover of $G'$. We note that all the construction that we did above is the same as the construction given in the proof of lemma $4.2$ of \cite{10.1007/978-3-031-52113-3_16}. So we can conclude, that $\mathcal{C}'$ is a sigma clique cover of $G'$ with $\operatorname{wgt}\left(\mathcal{C}'\right) \leq\left|V'\right|+\alpha-1$ and $G'$ does not contain any isolated vertices. The way we have defined $f$, it is clear that all cliques of $\mathcal{C}'$ are of the same size. 
\qed 

\begin{lemma}\label{equi sigma}
     $(G,k)$ is a yes-instance of {\sc UCIVS} if and only if $(G,|V|+k)$ is a yes-instance of {\sc sigma uniform clique cover}.
\end{lemma}
\proof Let $G_0, \ldots, G_{l}$ be a sequence of graphs with $G_0=G$ and $l \leq k$ such that each graph, except $G_0$, is obtained from its predecessor via a vertex split, and $G_{l}$ is a uniform cluster graph. By identifying all connected components of $G_{l}$ with their vertex sets, we can construct a Sigma uniform clique cover  $\mathcal{C}_{l}$ of $G_{l}$ with \texttt{wgt}$(\mathcal{C}_l)=|V(G_l)|$. Each split used in the construction of $G_0, \ldots, G_{l}$ introduces exactly one new vertex, therefore $|V(G_l)|=|V|+l$. Combining this with the fact that $l \leq k$, we derive \texttt{wgt}$(\mathcal{C}_l) \leq |V|+k$. Using the sequence $G_0, \ldots, G_{l}$ in reverse order, we iteratively apply \ref{merging} $l$ times using $\mathcal{C}_{l}$ and $G_{l}$ as base case and obtain $\mathcal{C}_0, \ldots, \mathcal{C}_{l}$. In particular, it follows that $\mathcal{C}_0$ is a Sigma uniform clique cover  of $G$ satisfying \texttt{wgt}$(\mathcal{C}_0)\leq |V|+k $  Thus, $(G,|V|+k)$ is a yes-instance.

For the reverse direction, let $\mathcal{C}$ be a Sigma uniform clique cover  of $G$ with $\texttt{wgt}(\mathcal{C}) \leq|V|+k$. By iteratively applying Lemma \ref{splitting} for a number of times, call it $l$, either until a uniform cluster graph is obtained as a direct result of the lemma, or alternatively, stopping after $l=k$ iterations, we can obtain a sequence of graphs $H_0, \ldots, H_{l}$ and their corresponding sigma uniform clique cover $\mathcal{C}_0, \ldots, \mathcal{C}_{l}$ respectively.

We shall now verify that also in the latter case where $l=k, H_{l}$ must be a uniform cluster graph. 
As a consequence of the $k$ applications of Lemma \ref{splitting}, we get $\texttt{wgt}(\mathcal{C}_{l}) \leq |V(H_{l})|$. 
By considering the fact that for each vertex $v \in V\left(H_{l}\right)$ there exists $C \in \mathcal{C}_{l}$ with $v \in C$ (since $\mathcal{C}_{l}$ is a sigma clique cover of $H_{l}$), we derive $ \mathrm{wgt} (\mathcal{C}_{l}) \geq |V(H_{l})|$.Thus, we have that $\texttt{wgt}(\mathcal{C}_l)=|V(H_{l})|$ and it follows that $\mathcal{C}_{l}$ forms a partition of $V\left(H_{l}\right)$. Using this partition property and the fact that $\mathcal{C}_{l}$ is a Sigma uniform clique cover  of $H_{l}$ allows us to directly conclude that $H_{l}$ is a uniform cluster graph. Thus, $H_{l}$ is a uniform cluster graph in both cases.

\qed

\begin{theorem}
    {\sc Uniform Cluster-Inclusive Vertex Splitting} parameterized by solution size admits a kernel with at most $4k$ vertices.
\end{theorem}

\proof Let $G_0,G_1,G_2,...G_l$ be a splitting sequence with size $l$ for $G$. Let $S = \{v_1,v_2,\ldots,v_l\}$ be the collection of vertices that were split to get $G_l$ from $G_0=G$. 

\begin{claim}
Let $v_i$ is split into $v_{i1}$ and $v_{i2}$. 
Set $S_0= \{v \in G : v \in N(v_{i1}) \cap N(v_{i2})\}$.
Then $S_0 \subseteq S$. 
\end{claim}
\proof For the sake of contradiction, let us assume that  $v \in S_0$ and $v \not\in S$. 
Since $v \in S_0$, there exists $v_i \in S$ such that $v \in N(v_{i1}) \cap N(v_{i2})$ for some $0 \leq i\leq l-1$. 
By application of the Lemma \ref{incident}, there are descendants of $v_{i1}$ and $v_{i2}$, say $v'_{i1}$ and $v'_{i2}$ respectively such that $v'_{i1}v \in G_l$ and $v'_{i2}v \in G_l$. 
Since $v_{i1}v_{i2} \not\in E(G_{j})$, there is no edge between $v'_{i1}$ and $v'_{i2}$ by Lemma \ref{noincident}. 
Hence $G[v'_{i1},v,v'_{i2}]$ forms an induced $P_3$ in $G_l$ which contradicts that $G_l$ is a cluster graph. 
Hence $S_0 \subseteq S$. \claimqed

\vspace{3mm}

Next, we observe that if there is indeed a split sequence of size at most $k$, then the vertices in  $V\setminus S$ have equal degrees in both $G$ and $G_l$. In other words, for every $x \in V\setminus S$, we have $d_{G}(x)=d_{G_l}(x)$. We assume that $G$ has at least $2k+1$ vertices; otherwise, we have a kernel of size $2k$.
We know that if the input is a yes-instance, then at least  $k+1$ vertices have the same degree, say $d$ and at most  $k$ vertices have degree not equal to $d$. 
Therefore, we first check if the input instance $(G,k)$ satisfies this condition. 
If the input instance does not satisfy this condition, then conclude that we are dealing with a no-instance. 
If the input instance satisfies this condition, we can calculate the exact value of $d$ in polynomial time. 
Assuming that the input is a yes-instance, let $c$ be the size of equal-sized cliques in $G_l$.
One can observe that $c$ must be equal to $d+1$ in  $G_l$.

Since the graph does not have isolated vertices, every vertex must be part of a clique in any Sigma Clique Cover.

\begin{ucivs}\label{UCIVS_combined}
    If any of the following conditions hold, conclude that we are dealing with a no-instance:
    \begin{enumerate}
        \item \( |\{v \in V(G) \mid d_G(v) > d\}| > k \),
        \item There exists a vertex \( v \in V(G) \) such that \( d_G(v) < d \),
        \item There exists a vertex \( v \in V(G) \) such that \( d_G(v) = d \) and \( G[N[v]] \) is not a clique.
    \end{enumerate}
\end{ucivs}

\noindent Note that the correctness of this Reduction Rule follows from the similar arguments in the proof of Lemma \ref{RR1 evs combined correct}.

\begin{lemma}\label{RR1 ivs combined correct}
    Reduction Rule \ref{UCIVS_combined} is correct.
\end{lemma}
\begin{proof}
We justify the correctness of each condition in Reduction Rule \ref{UCIVS_combined} as follows:

\begin{enumerate}
    \item \textbf{Condition 1}: If \( |\{v \in V(G) \mid d_G(v) > d\}| > k \), then there are at least \( k+1 \) vertices whose frequency must be at least two in any sigma $(d+1)$-clique cover. This means for any sigma $(d+1)$-clique cover of $G$ cost must be greater than $k$. Therefore, the instance must be a no-instance.
    

    \item \textbf{Condition 2}: If there exists a vertex \( v \in V(G) \) such that \( d_G(v) < d \), then clearly $G[N[v]]$ does not contain a clique of size $d+1$. It means $v$ can not be part of a clique in a sigma $(d+1)$-clique cover. Thus, the instance is a no-instance. 
     
    \item \textbf{Condition 3}: If there exists a vertex \( v \in V(G) \) such that \( d_G(v)=d \) and $G[N[v]]$ is not a clique. It means $G[N[v]]$ does not contain a clique of size $d+1$ and $v$ can not be part of a clique in a sigma $(d+1)$-clique Cover. Thus, the instance is a no-instance.
    
\end{enumerate}

Since all three conditions correctly identify invalid instances, the reduction rule is sound.  
\end{proof}

\noindent Now, we make two cases based on the values of $d$.

\noindent \textbf{Case 1}. Let us assume $d\geq 2k$

\noindent In this case, we provide a polynomial time algorithm to solve the problem, that is, due to Lemma \ref{equi sigma}, we determine whether there exists a sigma $(d+1)$-clique cover of $G$ with cost at most $k$.

\begin{tcolorbox}[colframe=black, colback=white, boxrule=0.5mm, title={Algorithm: Sigma $(d+1)$-Clique Cover}]
\begin{algorithmic}[1]\label{Sigma Cover}
\Require A graph \( G = (V, E) \) and an integer \( k \).
\Ensure A sigma $(d+1)$- clique cover \(\mathcal{C}\) of cost at most \( k \), or conclude that no such sigma cover exists.
\State Initialize an empty sigma cover: \(\mathcal{C} \gets \emptyset\)
\While{\( G \neq \emptyset \)}
    \State Find a vertex \( v \in V(G) \) such that \( d_G(v) = d \)
    \If{no such vertex exists}
        \State \Return "No instance."
    \EndIf
    \State Let \( C = G[N[v]] \), the subgraph induced by the closed neighborhood of \( v \)
    \If{\( C \) is not a clique}
        \State \Return "No instance."
    \EndIf
    \State Add \( C \) to \(\mathcal{C}\):
    \[
    \mathcal{C} \gets \mathcal{C} \cup C
    \]

    \State \textbf{Check Consistency of \(\mathcal{C}\):}
        Compute \(\text{cost}(\mathcal{C}) = \sum_{v \in V}  \max(0,\mathrm{freq}_{\mathcal{C}}(v)-1) \). If \(\text{cost}(\mathcal{C}) > k\), return "No instance."
    \State \textbf{Update \( G \) and \( k \):}
    \begin{align*}
    V(G) &\gets V(G) \setminus \{u \in N[v] \mid d_G(u) = d\}, \\
    k &\gets k - \mathrm{cost(\mathcal{C})}
    \end{align*}
\EndWhile
\State \Return \(\mathcal{C}\)
\end{algorithmic}
\end{tcolorbox}

\begin{lemma}\label{correctness of partition algorithm}
    The Sigma $(d+1)$-Clique Cover Algorithm (Algorithm~\ref{Sigma Cover}) correctly outputs a sigma $(d+1)$-clique cover of cost at most $k$, or conclude that no such clique cover exists.
\end{lemma}
\begin{proof}
The correctness of the algorithm is established as follows:

1. \textbf{Initial If Conditions}:  
   The first two \texttt{if} conditions (lines 4 and 7 of the algorithm) directly follow from \textbf{Reduction Rule \ref{UCIVS_combined}}:
   \begin{itemize}
       \item If no vertex \( v \) exists with \( d_G(v) = d \), then it is impossible to construct a valid $(d+1)$-sigma clique cover, as per \textbf{Condition 1} of the reduction rule.
       \item If \( G[N[v]] \) is not a clique for a vertex \( v \) with \( d_G(v) = d \), then \( v \) cannot belong to any \( K_{d+1} \)-clique, as per \textbf{Condition 3} of the reduction rule.
   \end{itemize}

2. \textbf{Cover Update}:  
   When a vertex \( v \) with \( d_G(v) = d \) is found, all edges adjacent to \( v \) must belong to the same part of the partition \(\mathcal{C}\). This is guaranteed by the definition of a $(d+1)$-sigma clique cover, which requires that \( C=G[N[v]] \) is a clique.  
   Therefore, adding \(C= G[N[v]] \) to \(\mathcal{C}\) is a valid and safe update.

3. \textbf{Consistency Check}:  
       The \textbf{consistency condition} holds due to \textbf{Lemma \ref{equi sigma}}, which guarantees that cost of the partition \(\mathrm{cost}(\mathcal{C})\) does not exceed the allowed budget \( k \).

4. \textbf{Graph Updates}:  
   Since all edges adjacent to the vertices in the set \(\{u \in N[v] \mid d_G(u) = d\}\) have been included in \(\mathcal{C}\) and any vertex in the set \(\{u \in N[v] \mid d_G(u) = d\}\) can be in only one clique, it is safe to delete these vertices. We delete edges $vu$ for all $u \in N(u)$ because edge $vu$ can be part of only one clique. If $w$ and $x$ are vertices with degree $d$ and $wx \in E(G)$, then we delete $wx$, because edge $wx$ can be part of exactly one clique. Additionally, the budget \( k \) is correctly reduced by the amount \(\mathrm{cost}(\mathcal{C})\).

5. \textbf{Termination}:  
   The algorithm terminates when \( G \) becomes empty. If any condition fails during execution, the algorithm correctly concludes a no-instance.

 Since $d\geq 2k$, every vertex of degree greater than $d$ is adjacent to a vertex of degree $d$ otherwise Reduction Rule UCIVS \ref{UCIVS_combined} could have been applied. Note that after each iteration, if $G$ is not empty, we will either have a vertex of degree $d$ or all vertices have a degree not equal to $d$. In the latter case, we return no-instance by Reduction Rule \ref{UCIVS_combined} and in the former case, since we have a vertex with degree $d$, we execute the line 3 of the algorithm.  This means that either $G$ becomes empty or any condition fails during execution.

Thus, the algorithm is correct.
\end{proof}

\noindent \textbf{Case 2}. Let us assume $d < 2k$.

There is a subtle difference between the two cases. In the case 1, after each iteration of updating \( G \), one of two outcomes occurs. Either the graph \( G \) becomes empty, or \( |V(G)| \geq 2k+1 \). In the former case, we can output the partition directly. In the latter case, Reduction Rule \ref{UCEVS_combined} guarantees that at least \( k+1 \) vertices (and all but \( k \)) have degree \( d \). This ensures that if the reduced input is a yes instance, the equal-sized cliques in the final graph must be of size \( d+1 \), which is consistent with the original instance.

In case 2, we must ensure that after each update of \( G \), either the graph becomes empty or it contains at least \( 2k+1 \) vertices. In this case, Reduction Rule \ref{UCIVS_combined} again ensures that at least \( k+1 \) vertices (and all but \( k \)) have degree \( d \). Since \( d \leq k \), the graph must have at least \( 4k+1 \) vertices to satisfy these conditions. Therefore, we apply the Sigma $(d+1)$-Clique Cover Algorithm (Algorithm~\ref{Edge Partition}) only if \( |V(G)| > 4k \).

\begin{ucevs}\label{UCEVS_d_leq_2k}
    If \( |V(G)| > 4k \), then run the Sigma $(d+1)$-Clique Cover Algorithm (Algorithm~\ref{Sigma Cover}) on the graph \( G \), but terminate the algorithm if the size of the vertex set becomes \( |V(G)| \leq 4k \).
\end{ucevs}

\noindent The correctness of the Reduction Rule \ref{UCEVS_d_leq_2k} follows from correctness of Sigma $(d+1)$-Clique Cover Algorithm (Algorithm~\ref{Sigma Cover}) due to Lemma \ref{equi sigma}.
This finishes the proof of Theorem \ref{UCIVSkernel}. \qed

\section{Kernelization algorithm for {\sc UCEE} parameterized by solution size}\label{Section UCEE}
In this section, we study the following problem: For a 
given graph $G$, can we transform $G$ into a uniform cluster graph by 
editing at most $k$ adjacencies, where editing involves adding or deleting   at most
$k$ edges? More formally, let $G=(V,E)$ be a graph. Then $F\subseteq V\times V$ is called
 a \emph{uniform cluster editing set} for $G$ if $G\triangle F$ is a uniform 
 cluster graph.
In this section, we prove the following theorem.

\begin{theorem}\label{UCEE}
    {\sc UCEE} parameterized by solution size admits a kernel of size $\mathcal{O}(k^2)$.
\end{theorem}

\noindent Note that many preprocessing rules applied in the case of {\sc UCEE}
are also applicable in the  cases of {\sc UCEA} and {\sc UCED}. To maintain conciseness, we will refer to the following preprocessing step as the {\it preparation step}, which begins here. This preprocessing step is the main idea in the kernelization algorithms for  {\sc UCEE}, {\sc UCED} and {\sc UCEA}.

\par A graph is a cluster graph
if and only if it does not have an induced $P_3$.  It is straightforward  to compute
a maximal set $\mathcal{P}_3$
of vertex-disjoint induced $P_3$s in $G$.
If $|\mathcal{P}_3|>k$, then we have a no-instance. 
Therefore, we assume that $|\mathcal{P}_3|\leq k$, and let $S$ be the vertices of $\mathcal{P}_3$. We have $|S|\leq 3k$. Let $\mathcal{C}$  denote the set of cliques of $G-S$.  Let $F$ be a uniform cluster editing set of size at most $k$ for $G$, and let 
 $V_F$ be the vertices of $F$. Then we have $|V_F|\leq 2k$.\\

We observe that if there indeed exists a uniform cluster editing set  $F$ of size at most $k$, then  the vertices in  $V\setminus V_{F}$ have equal degree in both $G$ and $G\triangle F$. 
In other words, for any two distinct vertices $x,y\in V\setminus V_{F}$, we have $d_{G}(x)=d_{G}(y)=d_{G\triangle F}(x)=d_{G\triangle F}(y)$. We assume that $G$ has at least $4k+1$ vertices; otherwise, we would have a kernel of size $4k$.
It is known  that if the input is a yes-instance, then  at least  $2k+1$ vertices
have the same degree $d$, and at most  $2k$ vertices  have degrees not equal to $d$.  
   Therefore, we first check if  the
input instance $(G,k)$ satisfies this condition. 
If the input instance does not satisfy this condition, then conclude that we are dealing 
with a no-instance.
If the input instance satisfies this condition, we can calculate the exact value of $d$ in polynomial time. Assuming that the input is a yes-instance, let $c$ be the size of equal sized cliques in $G\triangle F$.
It follows that $c$ must be equal to $d+1$ in  $G\triangle F$. 
This implies that the minimum degree of  $G$  is at least $d-k$, because  executing  $k$  edge editing operations can increase the degree of a vertex by at most  $k$.
Similarly, the maximum degree of $G$ is at most $d+k$, because  executing  $k$ edge editing operations can decreases the degree of a vertex by at most $k$. The preparation step ends here. \\

\noindent Next, we provide some Reduction Rules.

\begin{eee}\label{EEER0}
 If $\delta(G)<d-k$ or $\Delta(G)>d+k$ then conclude that we are dealing with  a no-instance.
\end{eee}

\begin{eee}\label{EEER01} If  the number of vertices in $V(G)$ with degrees
 not equal to $d$ exceeds $2k$, then conclude that we are dealing with a
 no-instance.
\end{eee}

\noindent We will consider two cases based on the value of $d$.
\subsection{\bf Case 1:} Let us assume that  $d\geq 6k$.
\begin{lemma}\label{c2k}
If $(G,k)$ is a yes-instance, then  for each $C \in \mathcal{C}$, we have $|C|>2k$.
\end{lemma}
\proof If there is a clique $C \in \mathcal{C}$ such that $|C|\leq 2k$, then every 
vertex $v\in C$ has a degree at most $5k-1$. This is because $v$ has at most $2k-1$ neighbours within $C$ and at most $3k$ neighbours in $S$. Given $d\geq 6k$, we have $5k-1<d-k$. This implies
$\delta(G)<d-k$.
Accordingly  to Reduction Rule \ref{EEER0}, if $\delta(G)<d-k$, then we conclude that we are dealing with  a no-instance. 
Therefore,  we must have $|C|>2k$. \qed \\

\begin{eee}\label{EEER1}
If $s\in S$ is adjacent to at least $k+1$ vertices of a clique $C \in \mathcal{C}$,
then add all missing edges between $s$ and $C$ and  decrement the parameter 
    $k$ by $|W|$, where $W$ is the set of all missing edges between $s$ and $C$. The resulting instance is $(G+ W,k-|W|)$. If $|W| \geq k+1$, then conclude that we are dealing with a  no-instance.  
\end{eee}

\begin{lemma}
    Reduction rule \ref{EEER1} is safe.
\end{lemma}
\proof The vertices of $C$ must belong to the same connected component in the edited graph. Since $s$ has at least $k+1$ neighbours in $C$, $s$ must be in the same connected component as the vertices in $C$. Otherwise, to place $s$ in a separate component,  we would need to delete at least $k+1$ edges between $s$ and $C$, which contradicts our assumption that $|F|\leq k$. Therefore, we should add all the missing edges between $s$ and $C$.\qed\\

\begin{eee}\label{EEER2}
If $s\in S$ is not adjacent to at least $k+1$ vertices of a clique $C \in \mathcal{C}$, then delete all edges between $s$ and $C$, and  decrement the parameter 
    $k$ by $|W|$, where $W$ is the set of edges between $s$ and $C$. The  resulting instance is $(G\triangle W,k-|W|)$. If $|W| \geq k+1$, then 
conclude that we are dealing with  a no-instance.
\end{eee}

\begin{lemma}
    Reduction rule \ref{EEER2} is safe.
\end{lemma}
\proof All the vertices of $C$ must belong to the same connected component in the edited graph. Note that $s$ cannot be in the same connected component as the vertices in $C$, as it would require adding at least $k+1$ edges. Therefore, we must delete all the edges between $s$ and $C$.\qed\\

\noindent Observe that exhaustive application of reductions UCEE \ref{EEER1} and UCEE \ref{EEER2} 
ensures that for each  $s\in S$ and each $C\in \mathcal{C}$, either $s$ is adjacent  to every vertex of $C$ or $s$
is adjacent to no vertex of $C$. 
\begin{eee}\label{EEER3}
If $s\in S$ has neighbours in two distinct cliques  $C_1,C_2 \in \mathcal{C}$, 
then conclude 
that we are dealing with a no-instance.
\end{eee}

\begin{lemma}
    Reduction rule \ref{EEER3} is safe.
\end{lemma}

\proof Due to Reduction rules UCEE \ref{EEER1} and UCEE \ref{EEER2}, $s$ must be adjacent to every vertex of $C_1$ and $C_2$. This implies that the graph contains at least $2k+1$ edge disjoint induced $P_3$s. Therefore the input instance is a no-instance.\qed

\vspace{5pt}

\begin{eee}\label{EEER4}
If $s_1$ and $s_2$ are two non-adjacent vertices of $S$ and have a neighbour in  $C\in \mathcal{C}$, then add the edge $(s_1,s_2)$ and decrement the parameter $k$ by 1. The resulting instance is $(G+(s_1,s_2),k-1)$.
\end{eee}

\begin{lemma}
    Reduction rule \ref{EEER4} is safe.
\end{lemma}

\proof Due to reductions UCEE \ref{EEER1} and UCEE \ref{EEER2}, $s_1$ and $s_2$ are adjacent to every vertex of $C$. Thus, $s_1$ and $s_2$ must be in the same connected component as the vertices in $C$ in the edited graph. 
Therefore, we must add the edge $(s_1,s_2)$ to the solution.\qed

\begin{eee}\label{EEER5}
If $s_1$ and $s_2$ are two adjacent vertices of $S$ and have neighbours in $C_{1}\in \mathcal{C}$ and $C_{2}\in \mathcal{C}$ respectively, then delete the edge $(s_1,s_2)$ and decrement the parameter $k$ by 1. The resulting instance is $(G-(s_1,s_2),k-1)$.
\end{eee}

\begin{lemma}
    Reduction rule \ref{EEER5} is safe.
\end{lemma}
\proof The vertices of $C_{1}$ and  the vertices of $C_{2}$ must be in  different connected components in the edited graph. Therefore, $s_{1}$ and $s_{2}$  must also be in  different connected components in the edited graph. Therefore, we must delete the edge  $(s_1,s_2)$ and add it to the solution. \qed

\begin{lemma}\label{lemma2k} If $(G,k)$ is a yes-instance and none of the reduction 
rules UCEE \ref{EEER0} to UCEE \ref{EEER5} are applicable to $G$, then $G$ is a disjoint union of cliques, each  of size  $d+1$.
\end{lemma}
\proof  When none of the reduction 
rules UCEE \ref{EEER0} to UCEE \ref{EEER5} are applicable to $G$, then  $G$ is clearly a disjoint union of cliques. Now, we show that these cliques are of size $d+1$.
If there exists a clique of size more than $d+1$,
then there would be more than $2k$ vertices whose degree is not equal to $d$.
Similarly, if there exists a clique of size less than $d+1$,
then there would be more than $2k$ vertices whose degree is not equal to $d$, because as shown in Lemma \ref{c2k}, all  cliques  $C\in \mathcal{C}$ have size at least $2k+1$.
This scenario contradicts Reduction Rule UCEE \ref{EEER01}.\qed \\
\noindent In Case 1, we obtain a kernel of size $\mathcal{O}(1)$.
\subsection{ Case 2:} Let us assume that $d \leq 6k-1$. 
\noindent We partition $\mathcal{C}=G-S$ into four parts as follows:
\begin{itemize}
    \item $\mathcal{C}_{<}= \{C \in \mathcal{C}: |C|<d+1\}$
\item $\mathcal{C}_{0,d+1}= \{C \in \mathcal{C}: |C|=d+1 \mbox{ and no vertex in } C \mbox { has a neighbour in } S\}$
\item $\mathcal{C}_{1,d+1}= \{C \in \mathcal{C}: |C|=d+1 \mbox{ and some vertex in } C \mbox { has a neighbour in } S\}$
\item $\mathcal{C}_{>}= \{C \in \mathcal{C}: |C|>d+1\}$
\end{itemize}


\begin{lemma}
If $(G,k)$ is a yes-instance and  reduction 
rule UCEE \ref{EEER01} is not applicable to $G$, then  $ \bigcup\limits_{C\in \mathcal{C}_{>}} |V(C)| \leq 2k$.
\end{lemma}
\proof Note that every vertex in $ \bigcup\limits_{C\in \mathcal{C}_{>}} V(C)$ 
has degree more than $d+1$. 
Due to  Reduction rule \ref{EEER01}, the number of vertices with degree
more than $d+1$ is bounded by $2k$.

\begin{eee}\label{EEER6}
    If $|\mathcal{C}_{0,d+1}|>2k$, then retain only $2k+1$ cliques from $\mathcal{C}_{0,d+1}$ and discard the rest.
\end{eee}
\begin{lemma}
     Reduction Rule \ref{EEER6} is safe.
\end{lemma}
\proof   An edge editing set of size at most $k$ can change degrees of at most $2k$ vertices
in $\mathcal{C}_{0,d+1}$. In other words, an edge editing set of size at most $k$ can 
affect at most $2k$ cliques  in  $\mathcal{C}_{0,d+1}$. There will still exist at least one clique of size $d+1$ in the final graph. 
Therefore, we can retain only  $2k+1$ cliques.\qed 
\vspace{5pt}
\begin{lemma}
     $|\mathcal{C}_{1,d+1}|\leq 2k$.
\end{lemma}
\proof Each clique $C\in \mathcal{C}_{1,d+1}$, by definition,  contains at least
one vertex that 
has a neighbour in $S$. This implies that every clique in $\mathcal{C}_{1,d+1}$ contains at least one vertex whose degree is more than $d$. Due to Reduction Rule UCEE \ref{EEER01}, there are at most $2k$ such vertices. Therefore, $|\mathcal{C}_{1,d+1}|\leq 2k$. \qed

\noindent Therefore, we get $\bigg| \bigcup\limits_{C\in \mathcal{C}_{d+1}} V(C)\bigg| \leq (4k+1)(6k-1)=24k^2+2k-1$, where $\mathcal{C}_{d+1}=\mathcal{C}_{0,d+1} \cup \mathcal{C}_{1,d+1}$.

\begin{lemma} If $(G,k)$ is a yes-instance and  reduction 
rule UCEE \ref{EEER0} is not applicable to $G$,  then we  have $ \bigg|\bigcup\limits_{C\in \mathcal{C}_{<}} V(C)\bigg| < 21k^{2}+5k$.
\end{lemma}
\proof Since Reduction Rule UCEE \ref{EEER0} is not applicable to
$G$, every vertex in $G$ has a degree at most $d+k\leq 7k-1$. 
For the sake of contradiction, assume that  $ \bigcup\limits_{C\in \mathcal{C}_{<}} |V(C)| \geq 21k^{2}+5k$. Each clique $C\in \mathcal{C}_<$ has size less than $d+1$, 
hence the
vertices in $C$ have degree less than $d$. As $(G,k)$ is a yes-instance, $G$ has a uniform cluster edge editing set of size at most $k$ that can increase the degree of at most $2k$ vertices. 
Therefore, at least $21k^{2}+5k-2k$ vertices in $ \bigcup\limits_{C\in \mathcal{C}_{<}} V(C)$ have neighbours in  $S$.  Since $|S|\leq 3k$ and $S$ has at least  $21k^{2}+3k$
neighbours in $G-S$, by the Pigeonhole principle, there is a  vertex $s\in S$ with  
at least $7k+1$ neighbours in $G-S$, which contradicts  the fact that $d_G(v)\leq 7k-1$ for all $v\in V$. 
Therefore, we conclude that $ \bigcup\limits_{C\in \mathcal{C}_{<}} |V(C)| < 21k^{2}+5k$. \qed \\

\noindent Applying the above reduction rules and results, we find that:
\begin{equation*}
\begin{split}
|V(G)| & = |S|+|V\setminus S|\\
 & = |S|+ \bigg|\bigcup\limits_{C\in \mathcal{C}_{<}} V(C) \bigg| +\bigg| \bigcup\limits_{C\in \mathcal{C}_{d+1}} V(C)\bigg|+\bigg|\bigcup\limits_{C\in \mathcal{C}_{>}} V(C) \bigg|\\
   & \leq 3k+(21k^{2}+5k)+(24k^2+2k-1)+2k\\
   &= 45k^{2}+12k-1. 
\end{split}          
\end{equation*}

\section{Parameterized Complexity of {\sc UCED} and {\sc UCEA}}
\label{Section UCED}

In this section, we see a number of parameterized complexity results regarding the complexity of both 
the {\sc UCED} and {\sc UCEA}.

\subsection{Kernelization Algorithm for \sc{UCED}}

\noindent In this section, we provide kernelization algorithms for {\sc UCED}.

\begin{theorem}\label{kernel-UCED}
   The {\sc UCED} problem parameterized by solution size admits a kernel with at most $6k$ vertices.
\end{theorem}

\noindent \textbf{Proof of Theorem \ref{kernel-UCED}.} We start with the same preparation step as designed above. However, in this preparation step, we apply edge deletion operations instead of edge editing operations.

\begin{eed}\label{EED1}
 If  $\delta(G)<d$ or $\Delta(G)>d+k$, then 
 conclude that we are dealing with a no-instance.
\end{eed}

\begin{eed}\label{EED2}
 If the number of vertices in $V(G)$ with degrees
 not equal to $d$ exceeds $2k$, then conclude that we are dealing with a
 no-instance. \end{eed}

\begin{eed}\label{EED3}
If there exists a vertex $u$ of degree $d$ in $G$  such that 
the subgraph induced by $N[u]$ is not a clique, then conclude that we are dealing with  a no-instance.
\end{eed}

\begin{lemma}\label{lemEED3}
    Reduction Rule \ref{EED3} is safe.
\end{lemma}
\proof Note that if the input is a yes-instance, then every vertex $u$ of degree $d$ must be part of a  clique of size $d+1$ after deleting the edges of the solution. 
Since the degree of 
$u$ is $d$, this can only happen if the closed neighborhood of $u$
(denoted $N[u]$) forms a clique.
 Therefore, if the subgraph induced by $N[u]$ is not a clique,  the instance is a no-instance. \qed 

\begin{eed}\label{EED4} 
Let $u \in V(G)$ be a vertex with degree  $d$. If $u$ has a neighbour $v$ with
a degree 
strictly greater than $d$, then 
delete all the edges $(v,w)\in E(G)$ such that $w$ is neither $u$  nor a neighbour of $u$. 
The set of edges deleted is denoted as $W$.
The resulting instance is $G=(V(G),k-|W|)$.
Also if $|W|\geq k+1$ then conclude that we are dealing with a no-instance.
\end{eed}


\begin{lemma}
    Reduction Rule \ref{EED4} is safe.
\end{lemma}
\proof As we have seen from the proof of Lemma \ref{lemEED3}, $N[v]$ must form a clique 
component in the final graph.
Therefore, we must delete the edges $(v,w)\in E(G)$ such that $v\in N[u]$ but $w \in V(G)\setminus N[u]$. 
It is also worth noting that the condition $|V(G)|\geq 4k+1$ ensures that the resulting graph,  obtained by deleting the edges in the solution, must be a cluster graph, where every clique is of size exactly $d+1$. \qed

\begin{eed}\label{EED5} If there are multiple  cliques of size $d+1$, 
then we retain only $x$ cliques of size $d+1$ where $x(d+1)\geq 2k+1$ and remove the remaining cliques of size $d+1$.
\end{eed}

\begin{lemma}
    Reduction Rule \ref{EED5} is safe.
\end{lemma}
\proof The condition $x(d+1)\geq 2k+1$ in Reduction Rule UCED \ref{EED5} ensures that at least $2k+1$ vertices have degree exactly $d$,  which in turn ensures that the size of cliques, after deleting at most $k$ edges, remains exactly $d+1$. Therefore deleting the extra cliques does not change the size of a solution. \qed 
\vspace{15pt}
 
\noindent We consider two cases based on the size of $d$.\\

\noindent \textbf{Case 1: $d\geq 2k$}

\begin{lemma}
After applying Reduction Rules \ref{EED1}-\ref{EED5} exhaustively, we obtain a uniform cluster graph.
\end{lemma}
\proof First we will show that every vertex in the reduced graph has degree $d$. For the sake of contradiction assume there exists a vertex $v$ with a degree of at least $d+1$. 
If $N(v)$ contains a vertex with degree $d$, then Reduction Rule UCED \ref{EED4} could have been applied.
Therefore, we assume that all vertices in $N(v)$ have a degree of at least $d+1$. 
As a result,  the number of vertices with degrees not equal to $d$ is at least $d+2\geq 2k+2$ as $d\geq 2k$.
However, in this scenario,  Reduction Rule UCED \ref{EED2} could have been applied
because the number of vertices with degrees not equal to $d$ exceeds $2k$. 
Consequently, in a reduced graph, every vertex has a degree exactly $d$.
After exhaustively applying Reduction Rules UCED \ref{EED3}, \ref{EED4} and \ref{EED5}, we obtain a clique of size $d+1$. This is a uniform cluster graph. \qed

\noindent Therefore in the case 1, we obtain a kernel of size $\mathcal{O}(1)$.
\vspace{5pt}

\noindent \textbf{Case 2: $d < 2k$}\\

\noindent After exhaustively applying Reduction Rules UCED \ref{EED3} and UCED \ref{EED4}, the components containing  a vertex of degree $d$ must form  a $d+1$-sized clique. After applying UCED \ref{EED5}, the cliques of size $d+1$ collectively contain at most $4k$ vertices. According to  Reduction Rule UCED \ref{EED2}, we have at most $2k$ vertices that do not have degree $d$. Therefore, we have $|V(G)| \leq 6k$. \\
   
\noindent This completes the proof of Theorem \ref{kernel-UCED}. 

\subsection{Kernelization Algorithm for \sc{UCEA}}

\noindent In this section, we provide kernelization algorithms for {\sc UCEA}.

\begin{theorem}\label{kernel-UCEA}
   The {\sc UCEA} problem admits a kernel with at most  $5k$ vertices.
\end{theorem}

\noindent \textbf{Proof of Theorem \ref{kernel-UCEA}.}  We start with the same preparation step. However, in this preparation step, we apply edge addition operations instead of edge editing operations.

\begin{eea}\label{EEA1}
 If  $\delta(G)<d-k$ or $\Delta(G)>d$, then 
 conclude that we are dealing with a no-instance.
\end{eea}

\begin{eea}\label{EEA2}
 If  the number of vertices in $V(G)$ with degrees
 not equal to $d$ exceeds $2k$, then conclude that we are dealing with a
 no-instance.\end{eea}

\begin{eea}\label{EEA3}
    If 
    there is a path between two nonadjacent vertices $u$ and $v$, then  make $u$ adjacent to $v$ and decrease $k$ by 1. The new instance becomes $(G+(u,v),k-1)$.
\end{eea}

\begin{lemma}
    Reduction Rule \ref{EEA3} is safe.
\end{lemma}
\proof Since there is a path between $u$ and $v$, they must belong to the same clique in the edited graph. Therefore, we must  connect $u$ and $v$ by an edge.\qed 
\vspace{5pt}
 
\noindent Observe that exhaustive application of  Reduction Rule UCEA \ref{EEA3} ensures that the new  graph is a collection of disjoint cliques.

\begin{eea}\label{EEA4}   If there are multiple  cliques of size $d+1$, 
then we retain only $x$ cliques of size $d+1$ where $x(d+1)\geq 2k+1$ and remove the remaining cliques of size $d+1$.

\end{eea}
\begin{lemma}
     Reduction Rule \ref{EEA4} is safe.
\end{lemma}
\proof Note that  Reduction Rule UCEA \ref{EEA4} ensures that at least $2k+1$ vertices have a degree exactly $d$, which in turn ensures that the size of cliques remains exactly $d+1$ after adding at most $k$ edges. Therefore,  deleting the extra cliques does not change the size of the solution. \qed 
\vspace{5pt}

\noindent \textbf{Case 1: $d\geq k+1$}

\begin{eea}\label{EEA5}
    If there is a clique of size less than $d+1$ in $G$, then  conclude that we
    are dealing with a no-instance.
\end{eea}

\proof Let us assume that we have a clique $C$ of size $x<d+1$.  To obtain a clique 
of size $d+1$  from $C$, we would need to add at least $x(d+1-x)$ edges. 
However, this exceeds the allowable number $k$ of edge addition operations, as 
 $x(d+1-x)>k$. \qed 
 \vspace{5pt}

\noindent Note that due to Reduction Rule UCEA \ref{EEA1}, there cannot be a clique of size more than $d+1$. Hence applying Reduction Rules UCEA \ref{EEA1} to UCEA \ref{EEA5}, all cliques are of size $d+1$. 
 
\noindent In Case 1, we obtain a kernel of size $\mathcal{O}(1)$.\\

\noindent \textbf{Case 2: $d\leq k$}\\

\noindent After applying Reduction Rule UCEA \ref{EEA4}, the cliques of size $d+1$ collectively contain at most $3k$ vertices. Accordingly to Reduction Rule \ref{EEA2}, there are at most $2k$ vertices with degrees not equal to $d$. Therefore, we have $|V(G)| \leq 5k$.
\vspace{5pt}

\noindent This completes the proof of Theorem \ref{kernel-UCEA}.

\subsection{An FPT Algorithm for UCED}

B$\ddot{o}$cker and Damaschke \cite{clusterfaster} provided an FPT algorithm for {\sc Cluster Edge Deletion}. This algorithm is based on branching technique. 
The overall idea is to provide a branching rule to destroy induced $P_{3}$ in the graph.
Once the branching rule is not applicable, the problem can be solved in polynomial time.
Note that as {\sc Uniform Cluster Edge Deletion} also requires destroying induced $P_{3}$'s, we can use the same branching rules in our case as well.
Once the branching rules are not applicable, we need to solve the problem in polynomial time.
For the sake of completeness, we explain the terminology and branching rule given by B$\ddot{o}$cker and Damaschke.
The score of an edge $e$ to be the number of induced $P_3$ that contains $e$. A graph is score-$s$ if every edge has score at most $s$. The score of a graph is the maximum score among its edges. Branching on an edge $e$ means to delete $e$, or to delete all edges that form induced $P_3$ with $e$.

{\it Doubling} a vertex $x$ of a graph means inserting a new vertex adjacent exactly to $x$ and its neighbors. The result of doubling several vertices does not depend on the order. Now we define several special $6$-vertex graphs, with the understanding that only the explicitly mentioned edges exist.
\begin{itemize}
    \item $3$-asterisk: a $K_3$ where each vertex is adjacent to one further vertex.
    \item $3$-sun: a $K_3$ where each two vertices are adjacent to one further vertex.
    \item fat $P_4$ : obtained from a $P_4$ by doubling both inner vertices.
    \item fat $P_5$: obtained from a $P_5$ by doubling its central vertex.
\end{itemize}

The disjoint union $G+H$ of graphs G and H consists of vertex-disjoint copies of $G$ and $H$, and $pG$ is the disjoint union of $p$ copies of $G$. The join $G*H$ is obtained from $G+H$ by inserting all possible edges between the vertices of $G$ and $H$.
Now, we state the theorem given by Damaschke \cite{score2}.

\begin{theorem}\label{score2} \cite{score2}
    The following graphs (with arbitrarily large positive $n, q, p$) and
their connected induced subgraphs comprise the complete list of connected score-$2$ graphs: $3$-asterisk, $3$-sun, fat $P_4$, fat $P_5; C_n (n \geq 4); K_q * C_5, K_q*K_3^c,
K_q*(K_2 + K_2); (qK_1 + pK_2)^c (p \geq  2)$.
\end{theorem}

\begin{lemma}\label{C4}
    If an induced $C_4$ exists, we can apply a 1.415 rule for Uniform Cluster Edge Deletion.
\end{lemma} 

\proof If we delete at most one edge from each pair of opposite edges in an induced $C_4$ then, obviously, some induced $P_3$ remains. Thus we must delete some pair of opposite edges, which yields the branching vector $(2,2)$. \qed

\vspace{3mm}

In \cite{clusterfaster} B$\ddot{o}$cker and Damaschke gave an FPT algorithm for cluster editing. The idea of the FPT algorithm was to first branch on the edge of a score larger than $2$. Then in the graph of score-$2$, use the classification given by the Theorem \ref{score2} and deal with each connected component separately. Our algorithm will also start with branching on the edge of a score larger than $2$. But for our problem, we will have to be careful about the size of each component. To deal with our problem, we will first guess the size of equal-sized cliques and deal with each component according to the guessed size. 

\begin{theorem}\label{FPT-UCED}
     {\sc Uniform Cluster Edge Deletion } is solvable in $\mathcal{O}(1.47^k)$ time.
\end{theorem}

\proof We know that deleting the vertices of any solution results in a disjoint union of equal-sized cliques (say, of size $c$). We start with guessing the size $c$. Note that the possible guesses for $c$ range from $1$ to $n$. If $c$ is a valid guess, then the number of vertices in $G$ must be a multiple of $c$.

As long as possible, take an edge $e$ of score larger than $2$, and delete $e$ or all edges forming a $P_3$ with $e$. The branching number is $1.47$. Once the graph is score-$2$, every component is one of the graphs from the Theorem \ref{score2}. Now we deal with each component to get a union of $c$-sized cliques. 

\begin{itemize}
    \item Note that the problem can be solved in polynomial time on  $3$-asterisk, $3$-sun, fat $P_4$, fat $P_5,b C_n (n \geq 4)$.
    \item In $(qK_1 + pK_2)^c$, we note that if $p \geq 2$, then we have an induced $C_4$. In this case, we can apply the $1.415$ branching rule in Lemma \ref{C4}. Therefore, we only need to deal with the case when $p=1$. If  $p=1$, then then the component is actually an induced subgraphs of $K_q*K_3^c$.
    \item The other graphs in the theorem consist of one clique $K$ of size $q$, joined with at most five other vertices. Suppose our connected component is $K_q * H$ where $|V(H)| \leq 5$. We note that $q+|V(H)|$ is a multiple of $c$. Otherwise we can return a no instance.
    
\vspace{5mm}    
    Suppose $F$ be our desired solution. Then $G-F$ must be disjoint union of c-sized cliques, say $C_1,C_2,\ldots,C_{\frac{q+|V(H)|}{c}}$.
    
    We guess a partition $\mathcal{P}= \{P_1,P_2,\ldots,P_l\}$ of vertices of $H$. We say a partition $\mathcal{P}= \{P_1,P_2,\ldots,P_l\}$ corresponds to a solution $F$ if $P_i=V(C_i) \cap V(H)$ and $G[P_i]$ is a clique. Otherwise $\mathcal{P}$ is a wrong guess. For each part $P_i$, there must exist $U_i \subseteq V(K_q)$ such that $G[P_i \cup U_i]=C_i$. We can easily find $U_i$ in polynomial time. This is because the vertices in $K_{q}$ are true twins.

 \vspace{5mm}   
    
    
    Note that $l \leq 5$. We take any arbitrary partition of $V(K_q)$, say $\mathcal{U}= \{U_1,U_2,...,U_r\}$ such that $|U_i|= c-|P_i|$ for $1\leq i \leq l$ and $|U_i|=c$ for $l+1 \leq i \leq \frac{q+|V(H)|}{c}$.
\end{itemize}

Define
$$
P'_i=
\begin{cases}
P_i \cup U_{i} \mbox{ for } 1 \leq i \leq l\\
U_{i} \mbox{ for } l+1 \leq i \leq \frac{q+|V(H)|}{c}
\end{cases}
$$

Now we have $\mathcal{P}'= \left\{P'_1,P'_2,\ldots,P'_{\frac{q+|V(H)|}{c}}\right\}$ a partition of vertices of the connected component. If for some $i$, $G[P'_i]$ is not a clique, we discard the partition $\mathcal{P}$. If every $G[P'_i]$ is a clique, we say that $\mathcal{P}$ is a \textit{valid} partition. For a valid partition, we find the number of edge deletions required to get the partition $\mathcal{P}'$ such that there is no edge between $G[P_i']$ and $G[P_i']$ for $i\neq j$. Suppose $k_\mathcal{P}$ denotes the number of edges deleted to get the partition $\mathcal{P}'$ for partition $\mathcal{P}$. Note that once we fixed $c$, the number $k_\mathcal{P}$ is also fixed.
For a guess $c$, We do the above process for all partitions of $V(H)$. Therefore, we solve the problem in polynomial time.

\qed

\subsection{A Subexponential algorithms for {\sc UCED} on Dense Graphs}

In this section, we will provide a subexponential algorithm for {\sc UCED} on everywhere $\alpha$-dense graphs. 
Let us begin by recalling the definition of everywhere $\alpha$-dense graphs.

\begin{definition} [\cite{DBLP:journals/jcss/AroraKK99,karpinski2008linear}]
    A graph on $n$ vertices is $\alpha$-dense if it has $\alpha n^2 / 2$ edges. 
    It is everywhere-$\alpha$-dense if the minimum degree is $\alpha n$. 
    We abbreviate $\Omega(1)$-dense as \emph{dense} and everywhere-$\Omega(1)$-dense as \emph{everywhere-dense}.
\end{definition}

We will use the subexponential algorithm of Lochet et al. for the \textsc{$d$-Way Cut} problem on everywhere $\alpha$-dense graphs as a subroutine in our algorithm. Specifically, they proved the following:
    
\begin{theorem}\label{d way cut}\cite{lochet_et_al:LIPIcs.STACS.2021.50}
    The \textsc{$d$-Way Cut} problem, parameterized by the size of the solution, admits an algorithm with running time 
    $ \left(\frac{1}{\alpha}\right)^{\mathcal{O}((1/\alpha)^3)} \cdot n^2 \cdot 2^{\mathcal{O} \left( \sqrt{\frac{k}{\alpha} \log \left( \frac{k}{\alpha} \right)} \right)} $
    on everywhere-$\alpha$-dense graphs.
\end{theorem}

We now state the main result of this section.

\begin{theorem}\label{UCED on dense graphs}
    The \textsc{UCED} problem, parameterized by the size of the solution, admits an algorithm with running time 
    $ \left(\frac{1}{\alpha}\right)^{\mathcal{O}((1/\alpha)^3)} \cdot n^2 \cdot 2^{\mathcal{O} \left( \sqrt{\frac{k}{\alpha} \log \left( \frac{k}{\alpha} \right)} \right)} $
    on everywhere $\alpha$-dense graphs.
\end{theorem}

\proof We will assume that $(\alpha n)^{2} \geq 49k$. If not then $n \leq \frac{7 \sqrt{k}}{\alpha}$ allowing us to try all possible vertex partitions and solve the problem in the required time.

Let the input instance be a yes-instance of \textsc{UCED}. 
Suppose that the graph \( G \) can be transformed into a uniform cluster graph by deleting a set \( F \subseteq E(G) \) of size at most \( k \), resulting in cliques of size \( h+1 \).
We begin by guessing \( h \), which has at most \( n \) possible values. Assume \( h \) has been correctly guessed. We distinguish between two cases:

\textbf{Case 1:} $h+1 < \alpha n -\sqrt{k}$.\\
Since every vertex \( v \in V(G) \) has degree \( d(v) \geq \alpha n \), it must be adjacent to at least \( \sqrt{k} \) edges in \( F \) to reduce its degree to \( h \). 
Thus, we must have \( \frac{n \sqrt{k}}{2} \leq k \), which implies \( n \leq 2\sqrt{k} \). In this case, we can solve the problem within the required time by exhaustively checking all vertex partitions.

\textbf{Case 2:} $h+1 \geq \alpha n -\sqrt{k}$.\\
Since \( \alpha n \geq 7 \sqrt{k} \), we conclude that \( h+1 \geq 6\sqrt{k} \). Let us denote by $L$ the set of vertices which are adjacent to at least $\sqrt{k}$ edges in $F$.
We have \( L = \{v \in V(G) \mid d(v) \geq h + \sqrt{k}\} \). Therefore \( |L| \leq 2\sqrt{k} \) because \( \frac{|L| \cdot \sqrt{k}}{2} \leq k \).

We have $G-F$ as a uniform cluster graph containing $\frac{n}{h+1}$ cliques of size exactly $h+1$.
Denote by \( C_1, C_2, \ldots, C_{\frac{n}{h+1}} \) the cliques in \( G - F\) with \( |C_i| = h+1 \) for all $1 \leq i \leq \frac{n}{h+1}$. 
Define \( C_i' = C_i \setminus L \).
Now, construct an instance \( I' = (G - L, d, k) \) of \textsc{$d$-Way Cut}, where \( d = \frac{n}{h+1} \). 
Since \( I \) is a yes-instance of \textsc{UCED}, \( I' \) is a yes-instance of \textsc{$d$-Way Cut}. 
To apply the algorithm in Theorem \ref{d way cut}, we need to make sure that the graph $G-L$ is everywhere dense.
We can assume that $ \alpha n-2\sqrt{k} > \frac{\alpha n}{2}$. 
Otherwise, we get $n \leq \frac{4 \sqrt{k}}{\alpha}$. 
In this case, we can try all the partitions and solve the problem in required time. 
Also, each vertex in $G$ has degree at least $\alpha n$. 
Therefore, the degree of every vertex in $G-L$ is at least $\alpha n-2\sqrt{k} \geq  \frac{\alpha n}{2}$.
Therefore, we can assume that $G-L$ is everywhere $\frac{\alpha}{2}$-dense.
Using the algorithm in Theorem \ref{d way cut}, we obtain a set of edges \( E' \subseteq E(G - L) \) of minimum size \( k' \leq k \) such that \( G - L - E' \) has exactly \( d \) connected components.

\begin{lemma}
    The \( d \) connected components obtained by deleting \( E' \) from \( G - L \) are precisely \( C_1', C_2', \ldots, C_d' \).
\end{lemma}
\proof Let \( A_1, A_2, \ldots, A_d \) denote the connected components obtained by removing \( E' \) from \( G - L \). 
Each \( C_i' \) satisfies \( |C_i'| \geq 4\sqrt{k} \) since \( |C_i| = h+1 \geq 6\sqrt{k} \) and \( |L| \leq 2\sqrt{k} \).
As we are deleting at most $k'\leq k$ edges in $G-L$, all but at most $\sqrt{k}$ vertices in $C'_{i}$ are in the some connected component $A_{j}$ for some $1\leq j \leq d$.
Suppose \( C_i' \) has all but at most \( \sqrt{k} \) of its vertices in a component \( A_j \).

\begin{claim}
    \( C_i' \subseteq A_j \)
\end{claim}
\proof Assume for contradiction that there exists \( x \in C_i' \) with \( x \in A_l \) for some \( l \neq j \). Construct a new partition \( B_1, B_2, \ldots, B_d \) by moving \( x \) from \( A_l \) to \( A_j \).
Let us denote by $E_{B}$ the set of edges needed to delete to get the connected components as $B_{1},\ldots,B_{d}$.
Note that $E_{B} =( E' \setminus \{ (x,w) ~|~ w \in A_{j} \})  \cup \{ (x,u) ~|~ u\in A_{l} \}. $
Therefore, $|E_{B}| = |E'| - d_{A_{j}}(x) + d_{A_{l}}(x)$.
Note that as $x\not\in L$, we know that $x$ has less than $\sqrt{k}$ neighbors outside $C_{i}$.
Therefore, we have $d_{A_{l}}(x)\leq \sqrt{k}$. 
We also know that $x$ has at least $3\sqrt{k}$ neighbors in $A_{i}$. Therefore, We have $d_{A_{j}}(x)\geq 3\sqrt{k}$.
This implies that $d_{A_{j}}(x) > d_{A_{l}}(x)$.
Hence, we get $|E_{B}|<|E'|$.
This is a contradiction to minimality of $E'$.
Therefore, we know that for every $i\in [d]$, there is an unique $j\in [d]$ such that $C'_{i} \subseteq A_{j}$. 
\qed \\

\noindent  Since all the connected components $A_{j}$'s are non-empty, for every $i\in [d]$, there is an unique $j\in [d]$ such that $C'_{i} = A_{j}$. This finishes the proof of the lemma. \qed \\

Due to the previous lemma, we exactly know the partition of all the vertices in graph $G$ except the vertices in $L$. But as we know that $|L|\leq 2\sqrt{k}$, we can guess for each vertex in $L$ its position in $A_{i}$'s. As $d = \frac{n}{h+1} \leq \frac{n}{\alpha n-\sqrt{k}} \leq \frac{n}{\big(\frac{\alpha}{2} \big) n} \leq \frac{2}{\alpha}$. This requires at most ${\big(\frac{2}{\alpha}\big)}^{\sqrt{k}}$ guesses. By trying all the possibilities, we can solve the problem in required time.  \\

\noindent  This finishes the proof of the Theorem \ref{UCED on dense graphs}. \qed

\section{Conclusion and Open Problems}
In this paper, we studied the problem of modifying a given graph so that the resulting graph becomes a collection of equal-sized cliques. We provided polynomial kernels for various types of modification problems along with FPT algorithms for few variants.
While we have made significant progress in understanding the parameterized complexity of the problems studied in this paper, several intriguing questions remain open for future research:

\begin{itemize}
    \item Does there exist a linear vertex kernel for {\sc UCVD}?
    \item Can we design a \((2 - \epsilon)^k \cdot n^{O(1)}\) algorithm for {\sc UCVD}?
    \item Is there a \(2^{O(k)} \cdot n^{O(1)}\) algorithm for {\sc UCEVS} and {\sc UCIVS}?
    \item Does a sub-exponential algorithm for {\sc UCEVS} and {\sc UCIVS} exist on everywhere dense graphs?
    \item Does there exist a linear vertex kernel for {\sc UCEE}?
    \item Construct an efficient FPT algorithm for {\sc UCEE} and {\sc UCEA}?
    \item Does a sub-exponential algorithm for {\sc UCEE} exists on everywhere dense graphs?
\end{itemize}

These questions represent exciting challenges in the parameterized complexity landscape and provide avenues for further exploration.

\bibliographystyle{abbrv}
\bibliography{references}
\end{document}